\documentclass[acmsmall]{acmart}

\AtBeginDocument{%
  }

\acmJournal{TOSEM} 

\usepackage[utf8]{inputenc} 
\usepackage{url}            
\usepackage{amsfonts}       
\usepackage{nicefrac}       
\usepackage{microtype}      

\usepackage{multirow}
\usepackage[table]{xcolor}  
\usepackage{subcaption}
\usepackage{algorithm}        
\usepackage{algorithmic}
\usepackage{amsthm}
\newtheorem{proposition}{Proposition}
\usepackage{enumitem} 
\usepackage{siunitx}
\usepackage{tabularx}

\makeatletter
\AtBeginDocument{%
  \fancypagestyle{firstpagestyle}{%
    \fancyhf{}%
    \renewcommand{\headrulewidth}{\z@}%
    \renewcommand{\footrulewidth}{\z@}%
    \fancyfoot[RO,LE]{\footnotesize \@journalNameShort}%
    \fancyhead[LE]{\ACM@linecountL\@acmBadgeL}%
    \fancyhead[LO]{\ACM@linecountL\@acmBadgeL}%
    \fancyhead[RO]{\@acmBadgeR}%
    \fancyhead[RE]{\@acmBadgeR}%
  }%
  \fancypagestyle{standardpagestyle}{%
    \fancyhf{}%
    \renewcommand{\headrulewidth}{\z@}%
    \renewcommand{\footrulewidth}{\z@}%
    \fancyfoot[RO,LE]{\footnotesize \@journalNameShort}%
    \fancyhead[LE]{\ACM@linecountL\@headfootfont\@acmArticlePage}%
    \fancyhead[RO]{\@headfootfont\@acmArticlePage}%
    \fancyhead[RE]{\@headfootfont\@shortauthors}%
    \fancyhead[LO]{\ACM@linecountL\@headfootfont\shorttitle}%
  }%
  \pagestyle{standardpagestyle}%
}
\makeatother

\acmJournal{TOSEM}
\acmVolume{ }
\acmNumber{ }
\acmArticle{ }
\acmYear{2026}
\acmDOI{}

\makeatletter
\renewcommand{\@formatdoi}[1]{%
  \ifx\@acmDOI\@empty
  \else
    \url{https://doi.org/#1}%
  \fi
}
\makeatother

\begin{document}

\title{SQL-o1: Automatic Text-to-SQL Program Synthesis via Self-Reward Heuristic Search}

\author{Shuai Lyu}
\email{Lxb_savior@bupt.edu.cn}
\affiliation{%
  \institution{Beijing University of Posts and Telecommunications}
  \country{China}
}
\authornote{Both authors contributed equally to this research.}

\author{Haoran Luo}
\authornotemark[1]
\email{haoran.luo@ntu.edu.sg}
\affiliation{%
  \institution{Nanyang Technological University}
  \country{Singapore}
}

\author{Ripeng Li}
\email{2966631798@bupt.edu.cn}
\affiliation{%
  \institution{Beijing University of Posts and Telecommunications}
  \country{China}
}

\author{Zhonghong Ou}
\email{zhonghong.ou@bupt.edu.cn}
\affiliation{%
  \institution{Beijing University of Posts and Telecommunications}
  \country{China}
}
\authornote{Co-corresponding authors.}




\author{Yifan Zhu}
\email{yifan_zhu@bupt.edu.cn}
\affiliation{%
  \institution{Beijing University of Posts and Telecommunications}
  \country{China}
}

\author{Tiesunlong Shen}
\email{tiesunlong@nus.edu.sg}
\affiliation{%
  \institution{National University of Singapore}
  \country{Singapore}
}
\authornotemark[2]


\begin{abstract}
Efficient human–database interaction is a core problem in software engineering. Text-to-SQL (Text2SQL) tackles this by mapping natural language questions to executable SQL queries, enabling automatic SQL synthesis and reliable interaction. Although large language models (LLMs) have advanced significantly, current approaches struggle with poor transferability to open-source LLMs, limited robustness against logic and function errors in complex queries, and inefficiencies in structured search. We introduce SQL-o1, a self-reward-driven heuristic search framework built on an agent-based architecture to enhance model reasoning capabilities. SQL-o1 leverages Monte Carlo Tree Search (MCTS) for structured, multi-step exploration and incorporates a dynamic pruning strategy to accelerate inference without sacrificing accuracy. On the Spider and BIRD benchmarks, SQL-o1 achieves a 10.8-point execution accuracy improvement on the complex BIRD dataset, surpassing even GPT-4-based models. It exhibits strong few-shot generalization and robust cross-model transferability across open-source LLMs. Our code is available at \href{https://github.com/ShuaiLyu0110/SQL-o1}{https://github.com/ShuaiLyu0110/SQL-o1}.

\end{abstract}




\maketitle

\section{Introduction}

As natural language interfaces become integral to modern software systems, the reliability of such interfaces has become a critical concern~\cite{Li_2024}. A key challenge lies in reliably translating natural-language requests into executable SQL programs for database-backed applications, a task commonly referred to as Text-to-SQL~\cite{10.1145/3746226}.

\begin{figure}[h]
    \centering
    \includegraphics[width=0.7\textwidth]{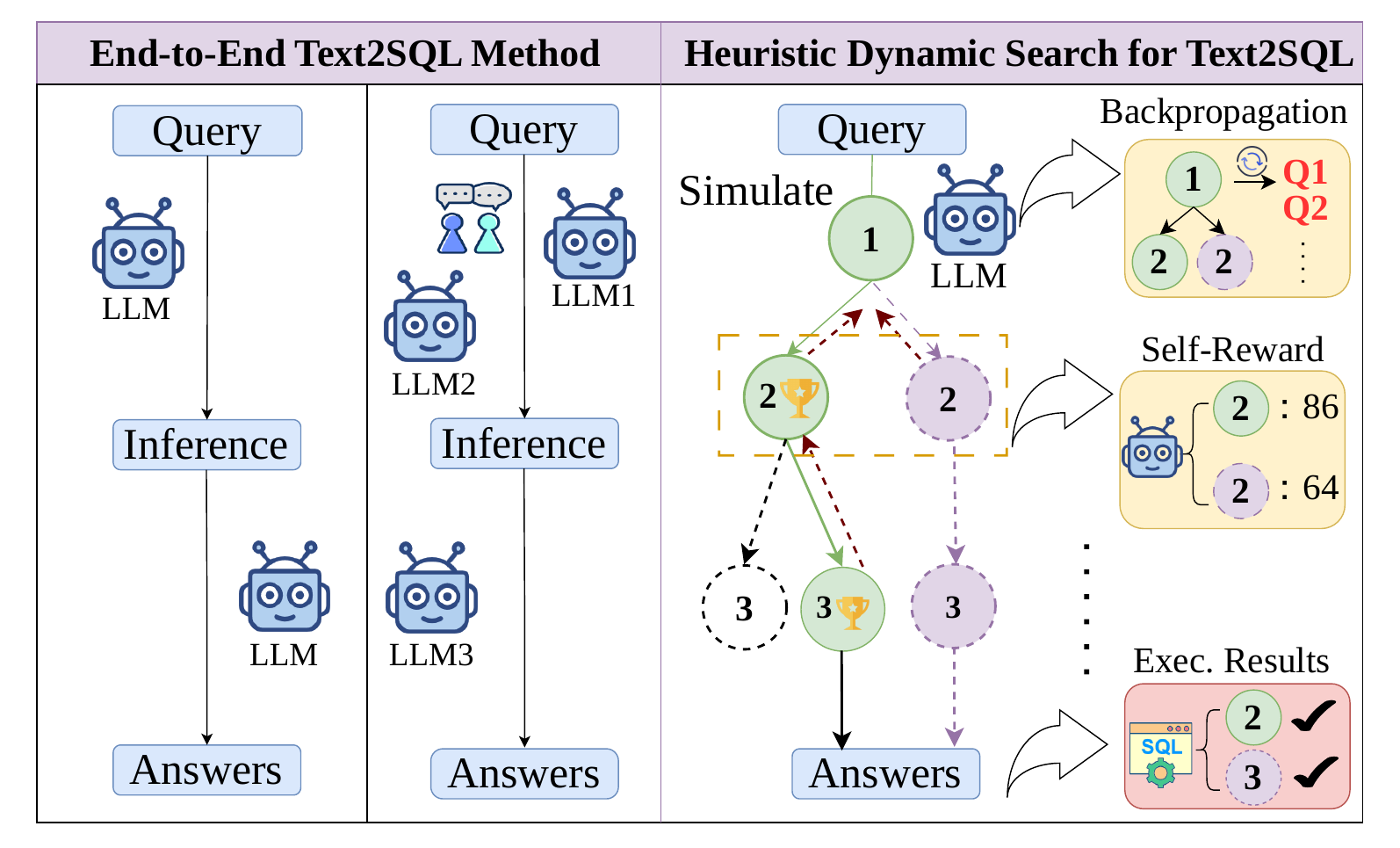}
    \caption{The figure illustrates the differences between end-to-end Text-to-SQL methods and heuristic dynamic search with self-reward. While end-to-end approaches directly map queries to answers via LLM inference, the proposed framework explores candidate paths with simulation, evaluates them using self-reward and execution results, and applies Back-propagation to guide more accurate SQL generation. }
    \label{fig:Intro}
    \Description[Comparison of Text2SQL approaches]{A diagram illustrating the key differences between end-to-end Text2SQL methods and the proposed Heuristic Dynamic Search with Self-Reward approach, highlighting the architectural and procedural distinctions.}
    \vspace{-4mm}
\end{figure}
Research on Text-to-SQL has progressed from relation-aware schema encoding and structured parsing~\cite{DBLP:conf/acl/WangSLPR20}, toward unified sequence-to-sequence formulations~\cite{DBLP:conf/emnlp/XieW0ZSYWZYWZWL22}, and more recently to large language model (LLM)-based approaches~\cite{lyu2026multibranch, luo2025kbqao1, Lyu_2025_tsvc}, which have substantially reshaped the broader landscape of Text-to-SQL~\cite{DBLP:journals/fcsc/ZhangWDZC25}.

However, many LLM-based methods continue to frame SQL generation as a single-pass decoding problem (Figure~\ref{fig:Intro}), employing techniques such as self-correction or Chain-of-Thought reasoning (CoT)~\cite{DBLP:conf/emnlp/TaiCZ0023}. We argue that this formulation underutilizes the inherently structured nature of SQL. Instead, we propose that SQL generation should be viewed as a controlled reasoning process, where iterative feedback, intermediate decision-making, and search heuristics play a central role.

\begin{figure*}[t]
    \centering
    \includegraphics[width=1.0\linewidth]{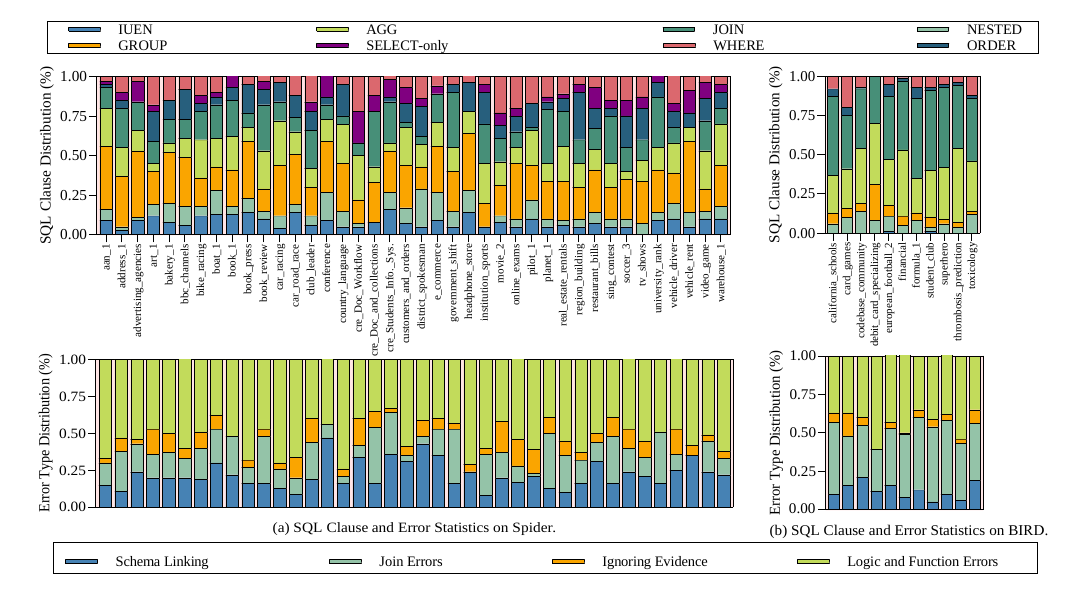}
    \caption{Distributions of SQL clauses and error types across databases in Spider and BIRD. While schema linking errors are common, logic and function errors are notably more frequent and persistent—highlighting the limitations of single-pass decoding in handling complex reasoning.}
    \label{fig:error_stats}
    \Description[Bar charts showing SQL clause distribution and error type frequency across databases]{Two side-by-side bar charts analyzing SQL characteristics and error patterns in Spider and BIRD datasets. The left chart shows the distribution of different SQL clauses (SELECT, WHERE, GROUP BY, etc.) across queries. The right chart compares the frequency of various error types, with schema linking errors being common but logic and function errors showing higher prevalence and persistence, demonstrating the challenges of complex reasoning in single-pass decoding approaches.}
    \vspace{-1\baselineskip}
\end{figure*}

To clarify the core challenges in Text-to-SQL, we identify three key limitations in current approaches: \textbf{(1) Poor transferability to open-source models.} Many state-of-the-art methods rely heavily on proprietary LLMs such as GPT-4, yet exhibit a sharp performance drop when deployed with open-source alternatives. \textbf{(2) Reasoning errors beyond single-pass decoding.} In addition to the well-known schema linking issues, we further analyzed Qwen2.5-7B after SFT on Spider~\cite{yu-etal-2018-spider} and BIRD~\cite{BIRD} by manually reviewing half of the errors, reclassifying them into four types, and then using GPT-4o with representative examples to categorize the rest (see Figure~\ref{fig:error_stats} and Table~\ref{tab:error_definition}). These errors persist even after fine-tuning, as single-pass decoding lacks the structured multi-step reasoning required to resolve them. \textbf{(3) Inefficiency in structured search.} Existing search-based approaches typically lack explicit process control or pruning strategies, resulting in expansive search spaces and inefficient inference for complex queries.

To address these challenges, we propose SQL-o1 (Figure~\ref{fig:Intro}), a structured generation framework that combines self-reward with heuristic search to improve Text-to-SQL generation. It adopts a general and extensible agent architecture with symbolic state-action interfaces, supporting its application across multiple open-source LLM backbones evaluated in this work. In particular, SQL-o1 introduces Progressive SQL Generation (PSG) to align cold-start fine-tuning with progressive SQL-prefix generation, thereby preparing the model for subsequent structured search. It further employs a confidence-based dynamic pruning strategy that adaptively controls the MCTS search space, improving overall inference efficiency while preserving promising SQL paths. Experiments on the Spider and BIRD benchmarks show that SQL-o1 consistently outperforms competitive open-source baselines and even surpasses GPT-4-based methods on particularly complex queries. A confidence-based pruning strategy improves inference efficiency without sacrificing accuracy, making SQL-o1 a scalable and plug-and-play solution for real-world applications.

\begin{table*}[t]
\centering
\caption{Design-space comparison between SQL-o1 and Text-to-SQL reasoning and search methods.}
\label{tab:design_comparison}

\fontsize{8.3}{9.2}\selectfont
\setlength{\tabcolsep}{2.5pt}
\renewcommand{\arraystretch}{1.05}

\begin{tabularx}{\textwidth}{
@{}
>{\raggedright\arraybackslash}p{2.1cm}
>{\centering\arraybackslash}p{2.2cm}
>{\centering\arraybackslash}X
>{\centering\arraybackslash}X
>{\centering\arraybackslash}p{1.3cm}
@{}
}
\hline
\textbf{Method}
& \textbf{Optimization}
& \textbf{Search / Inference Granularity}
& \textbf{Reward / Feedback}
& \textbf{Adaptive Pruning} \\
\hline

Reasoning-SQL~\cite{pourreza2025reasoning}
& GRPO
& Sequence-level reasoning
& Partial rewards
& No \\

SQL-R1~\cite{ma2025sql}
& SFT + GRPO
& Sequence-level reasoning
& Multi-component rewards
& No \\

CSC-SQL~\cite{sheng2025csc}
& GRPO
& Complete-SQL sampling/revision
& Execution feedback
& No \\

MCTS-SQL~\cite{yuan2025mctssql}
& Inference-time MCTS
& Complete-SQL refinement
& Refinement feedback
& No \\

Alpha-SQL~\cite{alpha-sql}
& Zero-shot MCTS
& High-level SQL actions
& Execution consistency
& No \\

\rowcolor{cyan!15}
\textbf{SQL-o1}
& \textbf{PSG-based SFT+MCTS}
& \textbf{Partial SQL prefixes}
& \textbf{Self-reward+online agreement}
& \textbf{Yes} \\
\hline
\end{tabularx}
\end{table*}

\section{Related Work}

\paragraph{Prompt Engineering.} In the era of large models, Text-to-SQL methods increasingly relied heavily on prompt engineering to enhance the reasoning capabilities of large language models (LLMs)~\cite{DBLP:conf/nips/PourrezaR23,DBLP:journals/corr/abs-2307-07306}. Techniques such as Chain-of-Thought (CoT)~\cite{DBLP:conf/emnlp/ZhangCCX023} and schema linking~\cite{wang2024macsql,DBLP:journals/pacmmod/LiZLFZZWP0024} have improved logical alignment with underlying databases. DAIL-SQL~\cite{dail_sql} provides a systematic analysis of prompts. Recent approaches have further adopted multi-step pipelines, integrating schema linking, self-correction, and execution feedback to iteratively optimize SQL generation~\cite{pourreza2024chase,xie2025opensearch,gao2024xiyan,DBLP:CHESS-sQL}. Additionally, some methods fine-tune open-source LLMs directly on large-scale Text-to-SQL corpora~\cite{DBLP:journals/pacmmod/LiZLFZZWP0024, DBLP:conf/emnlp/PourrezaR24}. DTS-SQL~\cite{DBLP:conf/emnlp/PourrezaR24} introduces a two-stage fine-tuning method that separates schema linking and SQL generation, significantly improving the performance of small LLMs. Building on this trend of improving model performance via fine-tuning, methods like SENSE~\cite{yang-etal-2024-synthesizing} and ROUTE~\cite{qin2024route} combine synthetic data with multi-task optimization strategies.

\paragraph{Agent-based Interaction with LLMs.}
With the capabilities of GPT-based large language models approaching human-level performance, there has been a growing interest in developing autonomous agents built on top of LLMs, particularly in the field of Text-to-SQL~\cite{DBLP:conf/iclr/ChenLSZ24,xiong2024interactive}. MAC-SQL~\cite{wang2024macsql} includes a refiner agent that uses SQL execution errors to correct and regenerate queries.
More closely related to our work, MCTS-SQL~\cite{yuan2025mctssql} first generates a complete SQL query and then employs MCTS to iteratively refine it, whereas SQL-o1 directly searches over progressively generated partial SQL prefixes after PSG-based cold-start fine-tuning. Alpha-SQL~\cite{alpha-sql} formulates zero-shot Text-to-SQL as an MCTS process with dynamically generated high-level SQL-construction actions and estimates terminal rewards through additional execution-consistency sampling. In contrast, SQL-o1 combines PSG-trained prefix completion with online execution agreement among regular MCTS rollouts and confidence-based dynamic pruning.
Recently, reinforcement learning has been increasingly introduced into Text-to-SQL tasks~\cite{ma2025sql,sheng2025csc,zhang2025reward}, yielding promising improvements by combining intermediate reward signals with prompt engineering techniques. While Reasoning-SQL~\cite{pourreza2025reasoning} makes a valuable contribution by designing task-specific rewards and leveraging the GRPO algorithm for policy optimization, SQL-o1 instead employs MCTS for step-wise SQL-prefix search without reinforcement-learning-based policy optimization. A more detailed design-space comparison with closely related Text-to-SQL methods is provided in Table~\ref{tab:design_comparison}.

\begin{table*}[t]
  \centering
  \caption{Distribution and explanation of defined error types across benchmarks.}
  \label{tab:error_definition}

  \small
  \setlength{\tabcolsep}{6pt}
  \renewcommand{\arraystretch}{1.10}

  \begin{tabularx}{\textwidth}{@{}lccX@{}}
    \toprule
    \textbf{Error Type}
    & \textbf{Spider (\%)}
    & \textbf{BIRD (\%)}
    & \textbf{Description} \\
    \midrule

    Schema Linking
    & 26.50
    & 12.8
    & Wrong or missing match between question and schema. \\

    Join Errors
    & 16.8
    & 39.7
    & Incorrect or missing join conditions between tables. \\

    Ignoring Evidence
    & 7.8
    & 5.8
    & Missing key information or constraints from the question. \\

    Logic and Function Errors
    & 48.9
    & 41.7
    & Logical or functional mistakes in query construction. \\

    \bottomrule
  \end{tabularx}

  \Description[Summary of error type distribution]{
  A table showing the percentage distribution of four types of errors
  (Schema Linking, Join Errors, Ignoring Evidence, Logic and Function Errors)
  across Spider and BIRD benchmarks, along with brief descriptions of each
  error type.}
  \vspace{-2mm}
\end{table*}

\section{Preliminaries}

\begin{definition}[Relational Database]
A relational database $\mathcal{D}$ consists of a set of tables $\mathcal{T} = {T_1, T_2, \dots, T_n}$, where each table is defined by a schema $\mathcal{C}_i = {c_1, c_2, \dots, c_m}$ containing column names and types. Tables may be interconnected through foreign keys, enabling multi-table queries.
\end{definition}

\begin{definition}[MCTS-based Search Tree]
The MCTS-based search tree $\mathcal{T} = (V, E)$ is a dynamic structure used to explore SQL decoding paths, where each node $v \in V$ corresponds to a decoding state $o_t$, and each edge $e \in E$ represents an action $a_t$ taken to expand the query from one state to the next. The tree is rooted at the initial state $o_0$, and is expanded through a sequence of MCTS rollouts, each consisting of selection, expansion, simulation, and back-propagation stages.
\end{definition}

\textbf{Problem Statement.}
Given a Text2SQL dataset $\mathcal{D}=\{(D_i,q_i,S_i)\}_{i=1}^{N}$, where each sample consists of an SQL database $D_i$, a natural language question $q_i$, and its corresponding ground-truth SQL query $S_i$, the goal of the Text2SQL task is to use a large language model to generate an SQL query $S_i'$ such that its execution result matches that of $S_i$.

\section{Method: SQL-o1}
\begin{figure}[h]  
    \centering
    \includegraphics[width=1.0\linewidth]{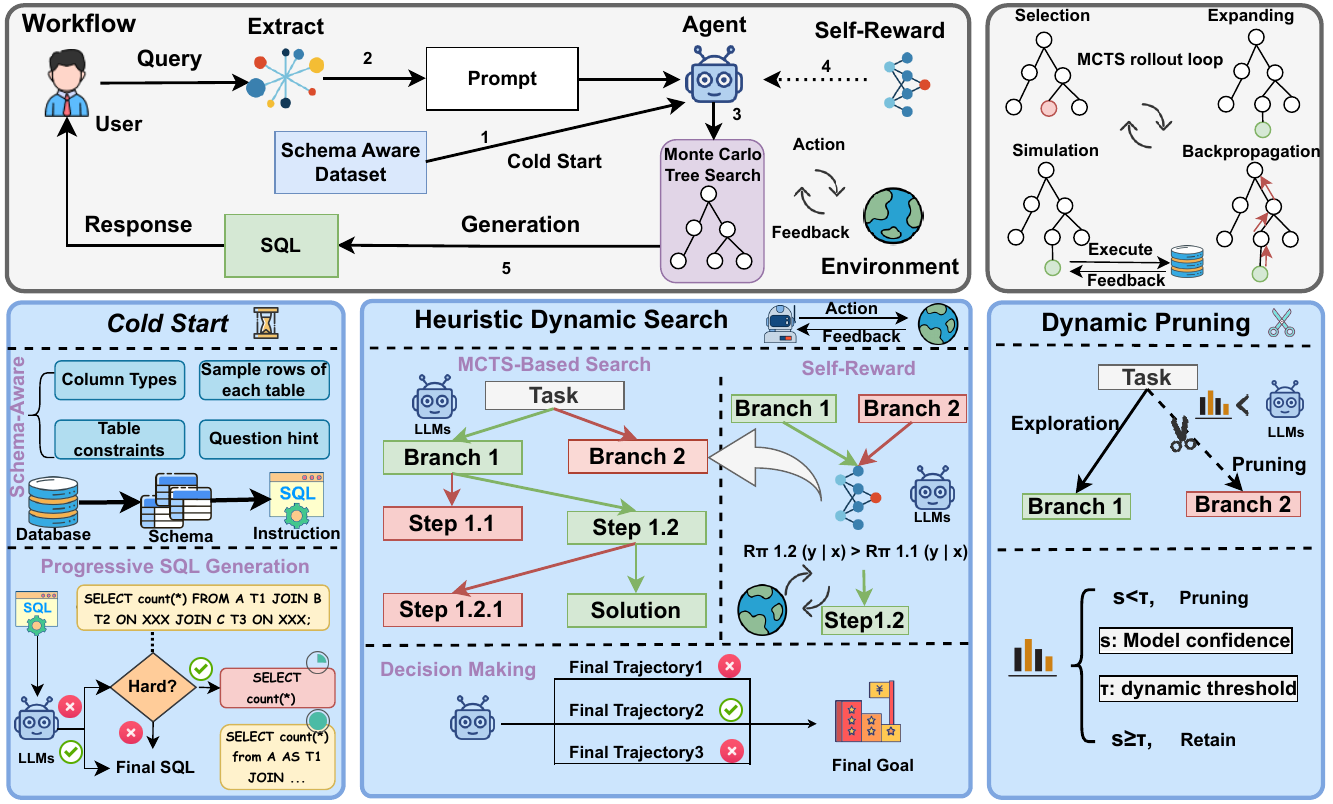}
    \caption{An overview of the implementation of the SQL-o1 framework.}
    \label{fig:framework}
    \Description[SQL-o1 framework implementation overview]{
        A diagram illustrating the complete implementation pipeline of the SQL-o1 framework, 
        showing the various processing stages from input to final SQL query output.
    }
    \vspace{-2mm}
\end{figure}

In this section, we introduce SQL-o1, a structured reasoning framework composed of three key components: (1) Agent Initialization, (2) Heuristic Dynamic Search with Self-Reward, and (3) Cold Start and Optimization, as illustrated in Figure~\ref{fig:framework}.

\subsection{Agent Initialization}
\label{sec:agent_init}
SQL-o1 initializes an autoregressive agent using schema-aware prompt, database environment, agent state space, and step-level exploration space. The agent incrementally constructs SQL queries by interacting with the environment and selecting valid actions, aiming to generate executable and semantically faithful SQL queries.
\paragraph{Agent Environment $\mathcal{E}$.}
The agent operates within a relational database environment \( \mathcal{E} \), which serves as the execution context and a source of feedback for evaluating candidate queries. Each database contains multiple tables with typed columns and relational constraints (e.g., primary and foreign keys). During inference, the agent grounds natural language questions to the database schema and incrementally constructs executable SQL queries through environment interaction.

Agent State Space $\mathcal{O}$. Let $\mathcal{O}=\{o_0,o_1,\ldots,o_t\}$ denote the sequence of intermediate SQL states, where $o_0$ is initialized from the natural language question $q$ and schema context $C$, and $o_t$ represents the SQL fragment constructed up to step $t$. At each step \( t \), the agent selects an action \( a_t \in \mathcal{A} \) by maximizing the conditional policy:
\begin{equation}
a_t = \arg\max_{a_t'} \pi(a_t' \mid o_{t-1}),
\label{eq:policy}
\end{equation}
where \( \pi \) is the policy model that scores candidate actions \( a_t' \) given the previous state \( o_{t-1} \).

Building on Section~\ref{sec:Schema_Preparation}, we collect failure cases where the pretrained model produces incorrect SQL predictions. From these, we generate prefix-truncated instances aligned with SQL syntax (e.g., \texttt{WHERE}, \texttt{AND}), forming partial inputs for completion. This yields a training dataset for Progressive SQL Generation (PSG), where the goal is to complete SQL queries from syntactically valid prefixes.
\begin{equation}
\mathcal{D}_\text{PSG} = \left\{ \text{Prefix}(S_i), S_i \right\}_{i=1}^{N}.
\end{equation}

This training paradigm enables the agent to reason over partial programs and promotes both compositional generalization and step-wise generation behavior from early training stages. PSG is used during cold-start fine-tuning to train the model on syntax-aligned SQL-prefix completion, thereby preparing it for subsequent MCTS search. It follows the canonical textual order of SQL to match the left-to-right decoding of autoregressive language models, rather than the logical execution order of a database engine. Since the natural-language question and complete database schema are provided before generation, the model has access to the tables and joins required by subsequent clauses when generating the \texttt{SELECT} clause. Therefore, this textual generation order does not conflict with the execution plan of the database engine.

\textbf{Agent Action Space \( \mathcal{A.} \)}
The action space \( \mathcal{A} \) defines the set of all tokens available to the agent during generation. It includes: (i) SQL syntax elements such as \texttt{SELECT}, \texttt{WHERE}, \texttt{GROUP BY}; (ii) column and table identifiers extracted from the schema \( D \); and (iii) literals and operators such as \textit{>}, \texttt{"USA"}, or \texttt{"2025"}. These components together enable grounded and compositional SQL generation.

\subsection{Heuristic Dynamic Search with Self-Reward}
\label{sec:Heuristic_search}
To guide SQL generation, we integrate Monte Carlo Tree Search (MCTS) with a self-reward mechanism within a search-based framework. This component balances exploration and exploitation to identify high-quality SQL candidates. It comprises two submodules: \textit{Self-Reward Evaluation}, which provides learning signals without explicit supervision, and \textit{Heuristic Dynamic Search}, which uses MCTS for structured, efficient query exploration.
\begin{figure}[t]  
    \centering
    \includegraphics[width=1.0\linewidth]{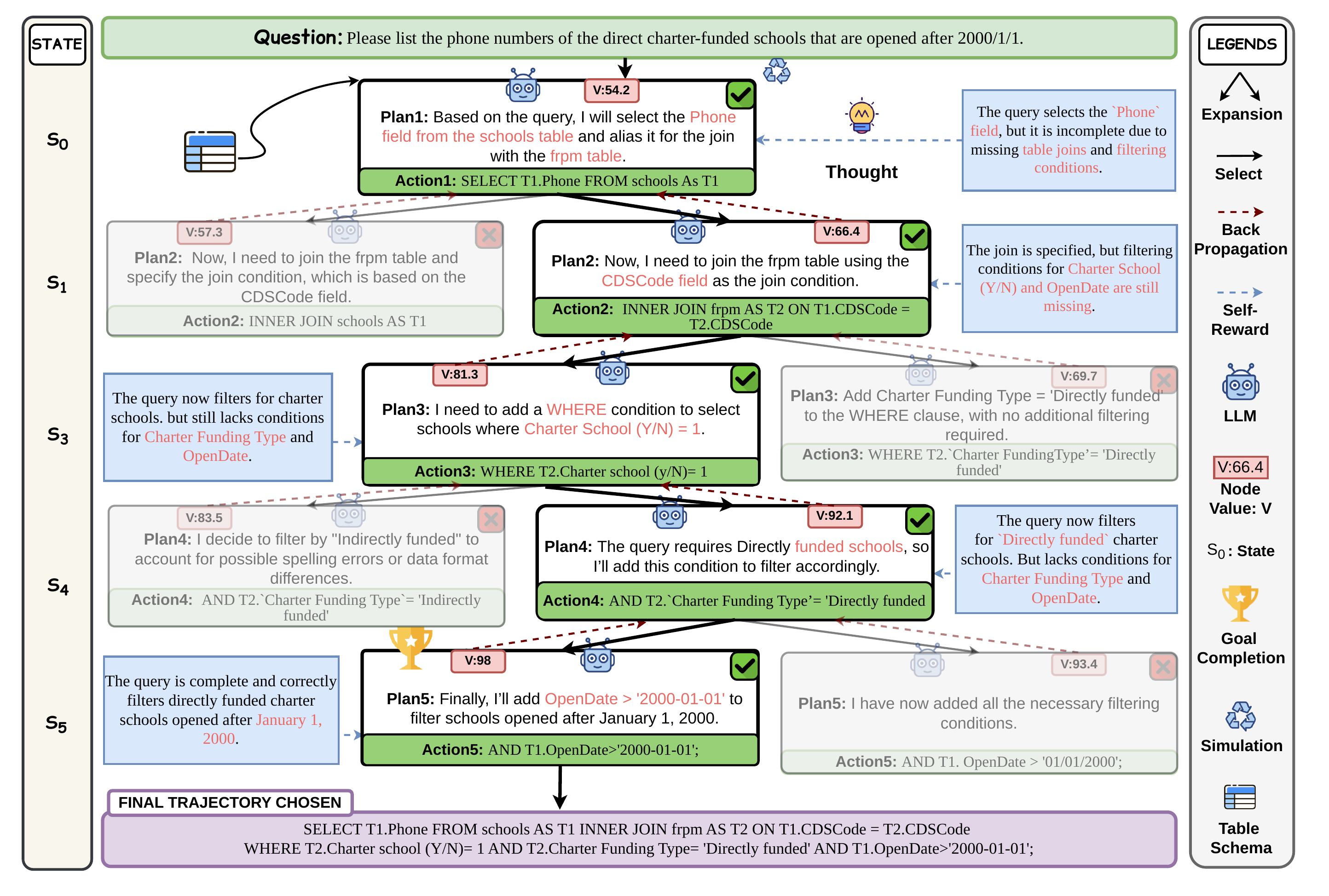}
    \caption{An example illustrates MCTS search guided by self-reward.}
    \label{fig:sql-o1}
    \Description[An example diagram of MCTS search process]{A detailed diagram showing the Monte Carlo Tree Search process guided by self-reward mechanism, illustrating the selection, expansion, simulation, and Back-propagation stages in SQL query generation. The diagram demonstrates how the algorithm explores different decoding paths to optimize the final SQL output.}
    \vspace{-2mm}
\end{figure}

\subsubsection{Self-Reward Evaluation}
To evaluate SQL fragment quality during decoding, we define a self-reward function \( R_{\pi} \) based on the agent’s output. The reward reflects the policy model $\pi$’s confidence via log-likelihood:
\begin{equation} 
R_{\pi}(y \mid x) = \beta + \alpha \log \pi(y \mid x),
\label{eq:reward}
\end{equation}
where $\beta$ is the base reward score (set to 100), and $\alpha$ (0.6) is a temperature controlling score variance.

\subsubsection{Heuristic Dynamic Search}
Monte Carlo Tree Search (MCTS) is a planning algorithm widely used in sequential decision-making~\cite{mcts}. In \texttt{SQL-o1}, we use MCTS as a structured heuristic to explore partial SQL programs via rollouts from intermediate states (Figure~\ref{fig}). Candidate actions are scored by self-reward, and value estimates are backpropagated to guide the overall search process toward executable, coherent queries.

\textbf{Selection.}
The selection phase starts from the root and traverses child nodes until reaching a leaf, where each node represents a partial SQL query. The next token is selected according to the policy in Equation~\eqref{eq:policy}, and heuristic truncation at semantic boundaries expands valid query fragments. Action selection at each level is guided by the upper confidence bound for trees (UCT)~\cite{mcts}:
\begin{equation}
n_t = \arg\max_{n\in\mathcal{N}(o_{t-1})} \left[ V(o_{t-1}+n) + w\sqrt{ \frac{\ln N(o_{t-1})} {N(o_{t-1}+n)} } \right],
\end{equation}
where $\mathcal{N}(\cdot)$ denotes the candidate expansions from the current state $o_{t-1}$, $V(\cdot)$ denotes the value estimate of an MCTS node, which is updated during back-propagation, and $N(\cdot)$ is the visit count of a given state.
\textbf{Expansion.}
When the selection process reaches a leaf node without children, the current SQL is incomplete, and the maximum depth \( L \) has not been exceeded, the agent invokes Beam Search to generate candidate SQL tokens:
\begin{equation}
\left\{ n_t^{(b)} \right\}_{b=1}^B \sim \pi \left( n_t \mid o_{t-1} \right)_{\text{Beam}},
\end{equation}
where \( \pi(\cdot)_{\text{Beam}} \) denotes beam decoding and \( B \) is the beam width. Candidate expansions are scored by the self-reward function \( R_\pi \), and the top-\( d \) fragments (\( d < B \)) are selected as child nodes:
\begin{equation}
\mathbb{N}(o_{t-1}) = \left\{ n_t^{(i)} \right\}_{i=1}^d = \underset{n \in n_t^{(b)} }{\arg\max}^d \ 
R_{\pi} \left( \left\{ n_t^{(b)} \right\}_{b=1}^B \mid o_{t-1} \right).
\end{equation}

For example, given a partial query over the \texttt{user} table, the agent may generate fragments like \texttt{SELECT user.id} or \texttt{SELECT user.name}, selecting the one most semantically aligned with the input question. The chosen fragment is then appended as a child node to the current leaf.

\textbf{Simulation and Back-propagation.}
After expansion, the agent performs simulated rollouts by sampling actions according to the policy $\pi$ (Equation~(1)). For the $n$-th rollout of length $l_n$, let $r_t^{(n)}=R_{\pi}(n_t^{(n)}\mid o_{t-1}^{(n)})$ denote the step-level self-reward at step $t$. We define the cumulative process reward as $R_{\mathrm{cum}}^{(n)} =\frac{1}{l_n}\sum_{t=1}^{l_n}r_t^{(n)}$. Once a terminal state is reached, the complete SQL query is evaluated using both model confidence and execution feedback from the environment. These signals are combined to compute the return of the current rollout:
\begin{equation}
\label{eq:Q_value}
G^{(n)} = \delta R_{\mathrm{cum}}^{(n)} + (1-\delta) \left[ R_{\pi}(S^{(n)}\mid q) + R_{\mathrm{exec}}(S^{(n)}) \right] , \qquad \delta\in(0,1).
\end{equation}
Here, $G^{(n)}$ denotes the return of the $n$-th rollout, $R_{\mathrm{cum}}^{(n)}$ summarizes the step-level self-rewards along the rollout, and $\delta$ balances the process-level and terminal reward signals, where \( R_{\text{exec}}(\cdot) \) denotes the online execution-agreement reward. During the \(n\)-th MCTS rollout, it is set to \(-1\) if SQL execution fails, \(+1\) if execution succeeds and the current non-empty normalized result has already been observed in at least one of the preceding \(n-1\) rollouts for the same question and database, and \(0\) otherwise. Although this reward does not incorporate ground-truth feedback, when combined with the model's relative preferences over candidate trajectories, it exploits repeated MCTS sampling to provide additional guidance for selecting repeatedly supported candidates.
The parameter $\delta \in (0,1)$ balances the process-level and terminal reward signals, with $\delta=0.5$ used in our experiments. Each rollout terminates when a complete SQL query is generated or when the maximum search depth is reached. The overall MCTS search terminates after completing the predefined \(N\) rollouts.
The algorithm then performs Back-propagation by updating the value estimates of all nodes along the trajectory from the leaf node to the root.
\begin{equation} 
 N(u)\leftarrow N(u)+1,\qquad V(u)\leftarrow V(u)+\frac{G^{(n)}-V(u)}{N(u)} .
\label{eq:Q_value_max}
\end{equation}
During back-propagation, the rollout return $G^{(n)}$ is propagated to each node $u$ along the selected trajectory. The visit count $N(u)$ is incremented, and the node value $V(u)$ is updated using the incremental sample mean in Equation~\eqref{eq:Q_value_max}.
\subsubsection{Query Trajectory Optimization}

To identify high-quality SQL candidates, the agent performs \( N \) rollouts using MCTS. For each rollout, it selects the trajectory ending in the highest-value state according to $V(\cdot)$ as the optimal generation path. The final SQL prediction \( \tilde{\mathcal{S}} \) is defined as:
\begin{equation} 
S\leftarrow\tilde{S}=\mathrm{MCTS}(q,\pi).
\end{equation}
where $\tilde{S}$ is the optimal SQL generated for the question $q$.

\textbf{Proposition 1.} \textit{Under the same base generation policy and the uniform positive-reachability condition, SQL-o1's self-reward-guided multi-rollout search has a strictly higher probability of generating at least one optimal SQL candidate than traditional single-pass generation when $N>1$.}
\begin{proof} 
We support this claim with results in Section~\ref{sec:main_result} and formal analysis in Section~\ref{appendix:proof}.
\end{proof}
\subsection{Cold Start and Optimization for SQL-o1 Reasoning}
To prevent instability and search drift when the agent decodes directly with MCTS, we apply cold-start training on a schema-aware dataset, embedding structural and semantic priors to guide search and accelerate convergence. At inference, confidence-based pruning removes low-probability candidates, reducing the search space and improving efficiency on complex queries.

\subsubsection{Data Preparation}
\label{sec:Schema_Preparation}
We construct a \textit{schema-aware training dataset} for cold-start supervised fine-tuning (SFT), embedding structural and semantic priors into the agent. This dataset includes column types (e.g., \textit{TEXT}, \textit{NUMBER}) for operator guidance, sampled cell values for literal prediction, and key constraints (e.g., primary/foreign keys) to ensure valid joins. This grounding improves execution awareness and step-wise decision-making in MCTS-based reasoning.
\textbf{Proposition 2.}  \textit{Given a fixed search depth $d$, a reduction in optimal-prefix prediction risk after PSG-based cold-start fine-tuning implies a non-decreasing expected log-likelihood of optimal SQL trajectories.}
\begin{proof}
Formal justification is provided in Section~\ref{appendix:proof2}, while the ablation results in Section~\ref{sec:ablation_study} provide downstream empirical evidence consistent with the strict risk-reduction case.
\end{proof}

\subsubsection{MCTS Dynamic Pruning}
To enhance rollout efficiency, we introduce a unified control mechanism inspired by DPTS~\cite{ding2025dynamic} that dynamically balances aggressive pruning and deep expansion. This strategy defines a step-wise confidence threshold \( \tau_t \) based on the model-assigned scores \( s_i \) over all candidate paths \( \mathcal{C} \) at step \( t \):
\begin{equation}
\tau_t = 
\begin{cases}
\lambda \cdot \frac{1}{|\mathcal{C}|} \sum\limits_{i \in \mathcal{C}} s_i, & \text{if } t \leq t_0 \\
\max\limits_{i \in \mathcal{C}} s_i, & \text{otherwise}
\end{cases}
\label{eq:tau_dynamic}
\end{equation}

Here, \( s_i \) denotes the model confidence for candidate path \( i \), \( \lambda \) is a scaling factor, and \( t_0 \) marks the transition from exploration to exploitation. During rollout, each candidate is compared against the dynamic threshold \( \tau_t \), and the control policy is applied:
\begin{itemize}
    \item If \( s_i < \tau_t \), the candidate path is pruned from further expansion.
    \item If \( s_i \geq \tau_t \), the candidate path is retained and expanded into deeper levels.
\end{itemize}

This formulation allows SQL-o1 to dynamically adjust search depth and breadth, allocating resources to promising subtrees while aggressively pruning unpromising ones.

\textbf{Proposition 3.} \textit{During MCTS rollouts, an optimal SQL prefix is retained under the confidence-based thresholding policy whenever the corresponding retention condition is satisfied.}
\begin{proof}
Empirical validation is presented in Section~\ref{sec:Inference_optim}, with theoretical analysis in Section~\ref{appendix:proof3}.
\end{proof}

\section{Theoretical Analysis}
\subsection{Proof of Proposition 1}
\label{appendix:proof}

\begin{proposition}
Under the same base generation policy and the uniform positive-reachability condition, SQL-o1's self-reward-guided multi-rollout search has a strictly higher probability of generating at least one optimal SQL candidate than traditional single-pass generation when $N>1$.
\end{proposition}

\textit{Proof.}

Let \(\mathcal{S}^{*}\) denote the set of optimal SQL queries, including semantically equivalent queries. Under the same base generation policy, traditional single-pass end-to-end generation produces one complete SQL candidate, denoted by \(\mathcal{S}^{(1)}\). Define its success probability as
\begin{equation}\label{eq:single_pass_success}
\mbox{\(\displaystyle
p_1
=
\Pr\!\left(
\mathcal{S}^{(1)}
\in
\mathcal{S}^{*}
\right),
\qquad
p_1\in[0,1)
\)}
\end{equation}
SQL-o1 instead generates \(N\) complete candidates, denoted by \(\mathcal{S}^{(1)},\ldots,\mathcal{S}^{(N)}\), through self-reward-guided MCTS rollouts. We assume that the first rollout follows the same base generation policy as traditional single-pass generation and therefore has success probability \(p_1\). For \(N\geq1\), define
\begin{equation}\label{eq:failure_probability_definition}
\mbox{\(\displaystyle
F_N
=
\bigcap_{n=1}^{N}
\left\{
\mathcal{S}^{(n)}
\notin
\mathcal{S}^{*}
\right\},
\qquad
P_N^{\mathrm{fail}}
=
\Pr(F_N)
\)}
\end{equation}
where \(F_N\) is the event that none of the first \(N\) rollouts generates an optimal SQL query, and \(P_N^{\mathrm{fail}}\) is the corresponding failure probability. We assume a uniform positive reachability condition: there exists \(\epsilon\in(0,1]\) such that, for every \(N\geq1\) with \(P_N^{\mathrm{fail}}>0\),
\begin{equation}\label{eq:positive_reachability}
\mbox{\(\displaystyle
\Pr\!\left(
\mathcal{S}^{(N+1)}
\in
\mathcal{S}^{*}
\mid
F_N
\right)
\geq
\epsilon
\)}
\end{equation}
This assumption imposes a uniform positive lower bound on the conditional probability of reaching an optimal SQL candidate and does not require different rollouts to be independent. If \(P_N^{\mathrm{fail}}=0\), an optimal candidate has already been generated with probability one. Otherwise, by the definition of \(F_N\) and the conditional probability rule,
\begin{equation}\label{eq:failure_contraction}
\mbox{\(\displaystyle
\begin{aligned}
P_{N+1}^{\mathrm{fail}}
&=
\Pr\!\left(
F_N
\cap
\left\{
\mathcal{S}^{(N+1)}
\notin
\mathcal{S}^{*}
\right\}
\right)\\
&=
P_N^{\mathrm{fail}}
\Pr\!\left(
\mathcal{S}^{(N+1)}
\notin
\mathcal{S}^{*}
\mid
F_N
\right)\\
&=
P_N^{\mathrm{fail}}
\left[
1-
\Pr\!\left(
\mathcal{S}^{(N+1)}
\in
\mathcal{S}^{*}
\mid
F_N
\right)
\right]\\
&\leq
(1-\epsilon)P_N^{\mathrm{fail}}.
\end{aligned}
\)}
\end{equation}
Since \(P_1^{\mathrm{fail}}=1-p_1\), iterating Equation~\eqref{eq:failure_contraction} gives
\begin{equation}\label{eq:failure_probability}
\mbox{\(\displaystyle
P_N^{\mathrm{fail}}
\leq
(1-p_1)(1-\epsilon)^{N-1}
\)}
\end{equation}
Therefore, the probability that SQL-o1 generates an optimal SQL query within \(N\) rollouts satisfies
\begin{equation}\label{eq:optimal_candidate_coverage}
\mbox{\(\displaystyle
\begin{aligned}
\Pr\!\left(
\exists n\leq N:
\mathcal{S}^{(n)}
\in
\mathcal{S}^{*}
\right)
&=
1-P_N^{\mathrm{fail}}\\
&\geq
1-(1-p_1)(1-\epsilon)^{N-1}.
\end{aligned}
\)}
\end{equation}
For \(N>1\), \(p_1<1\), and \(\epsilon>0\), the difference between this lower bound and the traditional single-pass success probability is
\begin{equation}\label{eq:strict_improvement}
\mbox{\(\displaystyle
\begin{aligned}
&1-(1-p_1)(1-\epsilon)^{N-1}-p_1\\
&\qquad=
(1-p_1)
\left[
1-(1-\epsilon)^{N-1}
\right]
>0.
\end{aligned}
\)}
\end{equation}
Hence, under the stated assumptions, SQL-o1's self-reward-guided multi-rollout search has a strictly higher probability of generating at least one optimal SQL candidate than traditional single-pass end-to-end generation. Equation~\eqref{eq:failure_probability} further shows that the upper bound on the failure probability decreases geometrically with the rollout budget \(N\).
\hfill\(\square\)

\subsection{Proof of Proposition 2}
\label{appendix:proof2}

\begin{proposition}
Given a search depth $d$, a reduction in optimal-prefix prediction risk after PSG-based cold-start fine-tuning implies a non-decreasing expected log-likelihood of optimal SQL trajectories.
\end{proposition}




\textit{Proof.}

Let

\begin{equation}
\tau^{*}
=
(a_1^{*},\ldots,a_d^{*})
\label{eq:appendix:optimal_trajectory}
\end{equation}

denote an optimal SQL trajectory of depth \(d\), and let

\begin{equation}
h_t^{*}
=
(s_0,a_{1:t-1}^{*})
\label{eq:appendix:optimal_prefix}
\end{equation}

denote its prefix state at step \(t\). PSG is used during cold-start fine-tuning to train the policy on syntax-aligned SQL-prefix completion. Let $\pi_0$ and $\pi_{\mathrm{psg}}$ denote the policies before and after PSG-based cold-start fine-tuning, respectively. For a policy $\pi$, define its expected prediction risk over optimal SQL-prefix states as
\begin{equation}
R_d(\pi)
=
\mathbb{E}_{\tau^\star}
\left[
\frac{1}{d}
\sum_{t=1}^{d}
-\log \pi(a_t^\star \mid h_t^\star)
\right].
\end{equation}

We treat risk reduction as an explicit condition rather than an automatic consequence of introducing PSG. Suppose that
\begin{equation}
R_d(\pi_{\mathrm{psg}})
\leq
R_d(\pi_0).
\label{eq:appendix:psg_risk_order} 
\end{equation}

The probability assigned by policy \(\pi\) to an optimal SQL trajectory is

\begin{equation}
P_{\pi}(\tau^{*}\mid s_0)
=
\prod_{t=1}^{d}
\pi(a_t^{*}\mid h_t^{*}).
\label{eq:appendix:optimal_path_prob}
\end{equation}

Taking logarithms and expectations gives

\begin{equation}
\mathbb{E}_{\tau^{*}}
\left[
\log
P_{\pi}(\tau^{*}\mid s_0)
\right]
=
-dR_d(\pi).
\label{eq:appendix:expected_log_likelihood}
\end{equation}

Combining Equations~\eqref{eq:appendix:psg_risk_order} and~\eqref{eq:appendix:expected_log_likelihood}, we obtain

\begin{equation}
\begin{aligned}
\mathbb{E}_{\tau^{*}}
\left[
\log
P_{\pi_{\mathrm{psg}}}
(\tau^{*}\mid s_0)
\right]
&
=
-dR_d(\pi_{\mathrm{psg}})
\\
&
\geq
-dR_d(\pi_0)
\\
&
=
\mathbb{E}_{\tau^{*}}
\left[
\log
P_{\pi_0}
(\tau^{*}\mid s_0)
\right].
\end{aligned}
\label{eq:appendix:psg_loglikelihood_order}
\end{equation}

Thus, under the risk-reduction condition, PSG-based cold-start fine-tuning does not decrease the expected log-likelihood assigned to optimal SQL trajectories. The inequality is strict whenever
\begin{equation}
R_d(\pi_{\mathrm{psg}})
<
R_d(\pi_0).
\end{equation}
The ablation results in Section~\ref{sec:ablation_study} provide downstream empirical evidence consistent with this strict case, rather than a direct measurement of the optimal-prefix prediction risk.

\hfill\(\square\)

\subsection{Proof of Proposition 3}
\label{appendix:proof3}
\begin{proposition}
During MCTS rollouts, an optimal SQL prefix is retained under the confidence-based thresholding policy whenever the corresponding retention condition is satisfied.
\end{proposition}

\textit{Proof.}
At decoding step \(t\), we use the ``optimal path'' to denote the current SQL prefix lying on an optimal complete SQL trajectory. If this prefix is retained, the corresponding optimal trajectory remains reachable in the MCTS tree.

At decoding step \( t \), the dynamic pruning threshold \( \tau_t \) is defined as:
\begin{equation}
\tau_t =
\begin{cases}
\lambda \cdot \frac{1}{|\mathcal{C}|} \sum\limits_{i \in \mathcal{C}} s_i, & \text{if } t \leq t_0 \\
\max\limits_{i \in \mathcal{C}} s_i, & \text{otherwise.}
\end{cases}
\end{equation}
where \( \mathcal{C} \) is the set of candidate paths at step \( t \), \( s_i \) is the confidence score of candidate \( i \), and \( \lambda \in (0, 1) \) is a scaling factor.

\textbf{Early-Stage Soft Thresholding (\( t \leq t_0 \)).} Let the optimal path have score \( s^* \), and the remaining candidates have scores \( s_2, \dots, s_n \):
\begin{equation}
\tau_t = \lambda \cdot \frac{1}{n} \sum_{i=1}^{n} s_i.
\end{equation}

To ensure the optimal path is not pruned:
\begin{align}
s^* &\geq \lambda \cdot \frac{1}{n} \left( s^* + \sum_{i=2}^n s_i \right) \notag \\
&= \frac{\lambda}{n} s^* + \frac{\lambda}{n} \sum_{i=2}^n s_i \notag \\
&s^* - \frac{\lambda}{n} s^* \geq \frac{\lambda}{n} \sum_{i=2}^n s_i \notag \\
&s^* \left( 1 - \frac{\lambda}{n} \right) \geq \frac{\lambda}{n} \sum_{i=2}^n s_i. \label{eq:prune_condition}
\end{align}

Define the average score of suboptimal candidates as:
\begin{equation}
\bar{r} = \frac{1}{n - 1} \sum_{i=2}^n s_i.
\end{equation}
\begin{equation}
\label{eq:early_retention_condition}
s^* \geq \frac{\lambda (n - 1)}{n - \lambda} \cdot \bar{r}.
\end{equation}

Equation~\eqref{eq:early_retention_condition} gives the exact retention condition for the optimal path in the early stage. Since \(\lambda<1\), the soft threshold allows the optimal path to be retained even when its confidence score is moderately lower than the average score of the remaining candidates.

\textbf{Late-Stage Hard Thresholding (\( t > t_0 \)).}
In this stage, the pruning threshold becomes strict:
\begin{equation}
\tau_t = \max_{i \in \mathcal{C}} s_i.
\end{equation}

In this case, the optimal path is retained if its confidence score is no lower than that of every competing candidate. Therefore, the corresponding retention condition is
\begin{equation}
s^* \geq \tau_t = \max_{i \in \mathcal{C}} s_i
\end{equation}

which shows that the late-stage policy retains the highest-confidence path without introducing any additional pruning criterion.

Let \( \lambda = 0.9 \), \( n = 5 \):

\begin{equation}
s^* \geq \frac{0.9 \cdot 4}{5 - 0.9} \cdot \bar{r}
= \frac{3.6}{4.1} \cdot \bar{r}
\approx
\mbox{\(0.88\)}
\cdot \bar{r}.
\end{equation}

Hence, if \( s^* \geq 0.88 \cdot \bar{r} \), the optimal path is not pruned in the early stage.

The two-phase confidence-based thresholding ensures that:
\begin{itemize}
    \item
    In early stages (\( t \leq t_0 \)), the optimal path is preserved whenever it satisfies the relaxed condition in Equation~\eqref{eq:early_retention_condition}.

    \item
    In later stages (\( t > t_0 \)), the optimal path is preserved whenever it has the highest confidence score among the current candidates.
\end{itemize}

Therefore, at each decoding step \(t\), the optimal SQL prefix is retained whenever the corresponding confidence condition is satisfied. Since retaining the optimal prefix preserves its continuation in the MCTS tree, if this condition holds at each decoding step, the complete optimal SQL trajectory remains reachable throughout the rollout.

\begin{table}[h]
  \centering
  \fontsize{6.5pt}{6.5pt}\selectfont
  \caption{We compare the performance of different models and methods on the Spider and BIRD datasets. To ensure fairness, we report ROUTE results only with open-source models of the same parameter scale. \textbf{Bold} numbers represent the best performance.}
  \resizebox{\textwidth}{!}{
  \Description[Comparison table of Text-to-SQL model performance on Spider and BIRD benchmarks]{
  This table presents a comprehensive comparison of various Text-to-SQL models and methods evaluated on the Spider and BIRD datasets. The table is organized into three main sections: Prompting with Closed-Source LLMs (primarily GPT-4-based methods), Prompting with Open-Source LLMs (including Llama3-based, Qwen2.5-based, and other methods), and Fine-Tuning with Open-Source LLMs (including Llama3-based, Qwen2.5-based, and other fine-tuned approaches). Performance is measured using execution accuracy (EX) and test suite accuracy (TS) metrics on the Spider development and test sets, and execution accuracy (EX) and valid efficiency score (VES) on the BIRD development set. The table clearly highlights the best performing results in bold across different model categories and parameter scales, demonstrating the effectiveness of the proposed SQL-o1 method compared to existing state-of-the-art approaches.
  }
  \label{table:1} 
  \begin{tabular}{lcccccc}
    \toprule
    \multirow{2}{*}{ \textbf{Model / Method}} & \multirow{2}{*}{ \textbf{Params}} & \multicolumn{3}{c}{\textbf{Spider}} & \multicolumn{2}{c}{\textbf{BIRD}} \\
    \cmidrule(lr){3-5} \cmidrule(lr){6-7}
     &  & Dev-EX & Dev-TS & Test-EX & Dev-EX & Dev-VES \\
    \midrule
    \multicolumn{7}{c}{\textit{\textbf{Prompting with Closed-Source LLMs}}} \\
    \midrule

    \textit{\textbf{GPT-4-based Methods}} & & & & & & \\
    GPT-4~\cite{achiam2023gpt} & - & 72.9 & 64.9 & - & 46.4 & 49.8 \\
    DIN-SQL~\cite{DBLP:conf/nips/PourrezaR23} & - & 82.8 & 74.2 & 85.3 & 50.7 & 58.8 \\
    DAIL-SQL~\cite{dail_sql} & - & 83.5 & 76.2 & 86.6 & 54.8 & 56.1 \\
    MAC-SQL~\cite{wang2024macsql} & - & 86.8 & - & 82.8 & 59.4 & 66.2 \\
    MCS-SQL~\cite{lee-etal-2025-mcs} & - & 89.5 & - & 89.6 & 63.4 & 64.8 \\
    \midrule

    \multicolumn{7}{c}{\textit{\textbf{Prompting with Open-Source LLMs}}} \\
    \midrule

    \textit{\textbf{Llama3-based Methods}} & & & & & & \\
    Zero-Shot~\cite{Llama3-8B} & 8B & 69.3 & 58.4 & 69.1 & 32.1 & 31.6 \\
    DIN-SQL & 8B & 48.7 & 39.3 & 47.4 & 20.4 & 24.6 \\
    MAC-SQL & 8B & 64.3 & 52.8 & 65.2 & 40.7 & 40.8 \\
    ROUTE + MCP & 8B & 75.0 & 63.4 & 72.0 & 42.7 & 44.8 \\
    \midrule

    \textit{\textbf{Qwen2.5-based Methods}} & & & & & & \\
    Zero-Shot~\cite{yang2024qwen2} & 7B & 72.5 & 64.0 & 75.9 & 41.1 & 42.0 \\
    DIN-SQL & 7B & 72.1 & 61.2 & 71.1 & 30.1 & 32.4 \\
    MAC-SQL & 7B & 71.7 & 61.9 & 72.9 & 46.7 & 49.8 \\
    ROUTE + MCP & 7B & 78.3 & 67.2 & 78.7 & 49.7 & 52.8 \\
    Zero-Shot-14B~\cite{yang2024qwen2} & 14B & 76.9 & 66.3 & 78.4 & 48.4 & 49.2 \\
    ROUTE + MCP & 14B & 80.0 & 67.3 & 80.6 & 56.3 & 57.6 \\
    \midrule

    \textit{\textbf{Others}} & & & & & & \\
    Mistral~\cite{jiang2023mistral} & 7B & 56.8 & 47.3 & 60.1 & 22.5 & 27.8 \\
    Gen. and Eval. Agents~\cite{DBLP:conf/acl/ChenWMP0024} + CodeLlama & 13B & 77.3 & - & - & 28.3 & - \\
    \midrule

    \multicolumn{7}{c}{\textit{\textbf{Fine-Tuning with Open-Source LLMs}}} \\
    \midrule

    \textit{\textbf{Llama3-based Methods}} & & & & & & \\
    SFT~\cite{Llama3-8B} & 8B & 82.4 & 76.2 & 83.1 & 53.1 & 59.0 \\
    \rowcolor{gray!15}
    ROUTE~\cite{qin2024route} & 8B & 86.0 & \textbf{80.3} & 83.9 & 57.3 & 60.1 \\
    \rowcolor{cyan!15} Ours: SQL-o1 & 8B & \textbf{87.4}$_{\pm 0.4}$ & 79.6$_{\pm 0.6}$ & \textbf{85.4}$_{\pm 0.4}$ & \textbf{63.4}$_{\pm 0.7}$ & \textbf{64.7}$_{\pm 0.8}$ \\
    \midrule

    \textit{\textbf{Qwen2.5-based Methods}} & & & & & & \\
    SFT~\cite{yang2024qwen2} & 7B & 80.9 & 75.6 & 82.8 & 51.4 & 53.1 \\
    \rowcolor{gray!15}
    ROUTE~\cite{qin2024route} & 7B & 83.6 & 77.5 & 83.7 & 55.9 & 57.4 \\
    \rowcolor{cyan!15} Ours: SQL-o1 & 7B & \textbf{84.7}$_{\pm 0.3}$ & \textbf{78.5}$_{\pm 0.5}$ & \textbf{85.1}$_{\pm 0.4}$ & \textbf{66.7}$_{\pm 0.6}$ & \textbf{70.4}$_{\pm 0.7}$ \\
    \midrule

    \textit{\textbf{Others}} & & & & & & \\
    DTS-SQL-7B~\cite{DBLP:conf/emnlp/PourrezaR24} & 7B & 82.7 & 78.4 & 82.8 & 55.8 & 60.3 \\
    CODES-7B + SFT~\cite{DBLP:journals/pacmmod/LiZLFZZWP0024} & 7B & 85.4 & 80.3 & - & 57.2 & 58.8 \\
    CODES-15B + SFT~\cite{DBLP:journals/pacmmod/LiZLFZZWP0024} & 15B & 84.9 & 79.4 & - & 58.5 & 56.7 \\
    SENSE-7B~\cite{yang-etal-2024-synthesizing} & 7B & 83.2 & 81.7 & 83.5 & 51.8 & - \\
    SENSE-13B~\cite{yang-etal-2024-synthesizing} & 13B & 84.1 & 83.5 & \textbf{86.6} & 55.5 & - \\
    \rowcolor{cyan!15} Ours: SQL-o1 + CodeLlama & 7B & 82.4$_{\pm 0.7}$ & 84.6$_{\pm 0.9}$ & 86.4$_{\pm 0.6}$ & 60.1$_{\pm 1.1}$ & 64.3$_{\pm 1.3}$ \\
    \rowcolor{cyan!15} Ours: SQL-o1 + DeepSeek-Coder & 7B & \textbf{87.0}$_{\pm 0.3}$ & \textbf{88.6}$_{\pm 0.4}$ & 86.5$_{\pm 0.3}$ & \textbf{64.5}$_{\pm 0.5}$ & \textbf{68.4}$_{\pm 0.6}$ \\
    \bottomrule
  \end{tabular}}
\end{table}

\section{Experiments}
\subsection{Experimental Settings}

\paragraph{Datasets.}
We evaluate SQL-o1 on Spider~\cite{yu-etal-2018-spider} and BIRD~\cite{BIRD}. Spider covers 200 databases across 138 domains with 7,000/1,034/2,147 splits, targeting cross-domain generalization. BIRD is more challenging, with 9,428/1,534/1,789 splits, richer schemas, and external knowledge requirements.
\paragraph{Baselines.}
We group the baselines into three categories: closed-source prompting, open-source prompting, and open-source fine-tuning. DIN-SQL~\cite{DBLP:conf/nips/PourrezaR23} decomposes Text-to-SQL into sub-problems; DAIL-SQL~\cite{dail_sql} selects demonstrations by question--query similarity; MAC-SQL~\cite{wang2024macsql} uses multiple agents for decomposition, schema retrieval, and SQL refinement; and MCS-SQL~\cite{lee-etal-2025-mcs} generates and selects multiple SQL candidates. For open-source prompting, we evaluate zero-shot decoding and ROUTE-MCP~\cite{qin2024route}. Fine-tuning baselines include SFT, ROUTE~\cite{qin2024route}, DTS-SQL~\cite{DBLP:conf/emnlp/PourrezaR24}, CODES~\cite{DBLP:journals/pacmmod/LiZLFZZWP0024}, and SENSE~\cite{yang-etal-2024-synthesizing}. Additional baseline details are provided in Section~\ref{sec:baseline_details}.

\paragraph{Evaluation Metrics.}
We evaluate Text-to-SQL performance using execution accuracy (EX), test suite accuracy (TS), and Valid Efficiency Score (VES). Let $M$ denote the number of evaluation examples, and let $S_i$ and $\hat{S}_i$ denote the reference and predicted SQL queries for the $i$-th example, respectively. EX measures the proportion of predictions whose execution results match those of the corresponding reference queries:
\begin{equation}
\mathrm{EX}
=
\frac{1}{M}
\sum_{i=1}^{M}
\mathbb{I}\!\left[
\mathrm{Exec}(\hat{S}_i,D_i)
=
\mathrm{Exec}(S_i,D_i)
\right].
\end{equation}
TS follows the Spider test-suite evaluation protocol, where a prediction is considered correct only if its execution result matches that of the reference query across all database instances in the corresponding distilled test suite. Formally, letting $\mathcal{T}_i$ denote the distilled test suite for example $i$,
\begin{equation}
\mathrm{TS}
=
\frac{1}{M}
\sum_{i=1}^{M}
\mathbb{I}\!\left[
\forall D\in\mathcal{T}_i,\,
\mathrm{Exec}(\hat{S}_i,D)
=
\mathrm{Exec}(S_i,D)
\right].
\end{equation}
For BIRD, we additionally report VES to jointly evaluate execution correctness and efficiency:
\begin{equation}
\mathrm{VES}
=
\frac{1}{M}
\sum_{i=1}^{M}
\mathbb{I}\!\left[
\mathrm{Exec}(\hat{S}_i,D_i)
=
\mathrm{Exec}(S_i,D_i)
\right]
\sqrt{\frac{E(S_i)}{E(\hat{S}_i)}},
\end{equation}
where $E(\cdot)$ denotes the execution efficiency of an SQL query under the BIRD evaluation protocol. Thus, incorrect predictions contribute zero to VES, while correct predictions are further weighted according to their relative execution efficiency. EX is reported for all datasets. TS is reported for Spider-based benchmarks when available, while VES is additionally reported for BIRD. TS is unavailable for Spider-DK and the Spider test set.

\begin{table}[hbt!]
\centering
\caption{Evaluation results of SQL-o1 and previous methods on Spider-based robustness benchmarks, including SYN, REALISTIC, and DK. TS results for DK are omitted due to compatibility issues.}
\label{table:2}
\begin{tabular}{lcccccc}
\hline
\textbf{Model/Method} & \textbf{SYN-EX} & \textbf{SYN-TS} & \textbf{REALI.-EX} & \textbf{REALI.-TS} & \textbf{DK-EX} & \textbf{AVG.} \\
\hline
SQL-PaLM (FS) & 74.6 & 67.4 & 77.6 & 72.4 & 66.5 & 71.7 \\
SQL-PaLM (FT) & 70.9 & 66.4 & 77.4 & 73.2 & 67.5 & 71.1 \\
SENSE-7B & 72.6 & 64.9 & \textbf{82.7} & \textbf{75.6} & 77.9 & 74.7 \\
\rowcolor{gray!15}
ROUTE + Llama3-8B & 77.4 & \textbf{70.2} & 80.9 & 72.6 & 74.6 & 75.1 \\
\rowcolor{cyan!15}
Ours: SQL-o1 + Llama3-8B & \textbf{77.6}$_{\pm 0.7}$ & 69.2$_{\pm 0.9}$ & \textbf{82.7}$_{\pm 0.6}$ & 72.8$_{\pm 0.8}$ & \textbf{78.7}$_{\pm 1.0}$ & \textbf{76.2}$_{\pm 0.6}$ \\
\hline
\end{tabular}
\Description[SQL-o1 robustness evaluation results]{
This table presents the evaluation results of the proposed SQL-o1 method and previous approaches on Spider-based robustness benchmarks, including SYN, REALISTIC, and DK (Database-Knowledge) test sets. Execution accuracy (EX) and Test Suite accuracy (TS) metrics are reported where available. TS results for the DK benchmark are omitted due to compatibility issues with the test suite framework. The SQL-o1 method with Llama3-8B backbone achieves the highest average performance (76.2\%) across all benchmarks, demonstrating superior robustness compared to existing methods including SQL-PaLM (both few-shot and fine-tuned variants), SENSE-7B, and ROUTE + Llama3-8B. Notably, SQL-o1 achieves the best performance on SYN-EX (77.6\%), REALISTIC-EX (82.7\%), and DK-EX (78.7\%) benchmarks, while maintaining competitive performance on the remaining metrics.
}
\end{table}

\subsection{Main Result}
\label{sec:main_result}
As shown in Tables~\ref{table:1} and~\ref{table:2}, prompt-based open-source models outperform zero-shot baselines but still lag behind closed-source counterparts due to scale and data limitations. Combined with open-source LLMs like Llama3-8B and Qwen2.5-7B, SQL-o1 achieves strong, domain-robust performance. On Spider and its variants, it matches ROUTE overall and outperforms it on complex queries via MCTS-based search. Notably, it reaches $66.7\%$ Dev-EX on BIRD—exceeding both ROUTE + Qwen2.5-7B ($55.9\%$) and GPT-4. Without extra training data, SQL-o1 yields consistent gains across models and tasks, showing robust generalization to diverse query complexities.

\begin{table}[htb!]
\centering
\begin{minipage}[t]{0.49\linewidth}
\vspace{0pt}
\centering
\caption{The ablation study of SQL-o1.}
\label{table:4}
\footnotesize
\setlength{\tabcolsep}{2.5pt}
\renewcommand{\arraystretch}{1.15}

\begin{tabularx}{\linewidth}{
@{}
>{\raggedright\arraybackslash}p{2.25cm}
*{3}{>{\centering\arraybackslash}X}
@{}
}
\toprule
\textbf{Model/Method}
& \textbf{Spider-EX}
& \textbf{Spider-TS}
& \textbf{BIRD-Dev} \\
\midrule
\multicolumn{4}{@{}l@{}}{\textit{Main Results Ablation Experiment}} \\
\midrule
\rowcolor{cyan!15}
SQL-o1
& \textbf{87.4}
& \textbf{79.6}
& \textbf{63.4} \\
w/o \textit{MCTS}
& $79.8_{\pm 0.8}$
& $74.3_{\pm 1.3}$
& $52.0_{\pm 0.6}$ \\
w/o \textit{PSG}
& $78.2_{\pm 1.1}$
& $73.4_{\pm 0.7}$
& $61.8_{\pm 0.5}$ \\
w/o \textit{Cold Start}
& $72.8_{\pm 0.7}$
& $62.2_{\pm 1.1}$
& $35.5_{\pm 1.4}$ \\
\bottomrule
\end{tabularx}
\end{minipage}
\hfill
\begin{minipage}[t]{0.49\linewidth}
\vspace{0pt}
\centering
\caption{Exploratory results of base models.}
\label{table:5}
\footnotesize
\setlength{\tabcolsep}{2.5pt}
\renewcommand{\arraystretch}{1.15}

\begin{tabularx}{\linewidth}{
@{}
>{\raggedright\arraybackslash}p{2.0cm}
*{3}{>{\centering\arraybackslash}X}
@{}
}
\toprule
\textbf{Model/Method}
& \textbf{Spider-Dev}
& \textbf{Spider-TS}
& \textbf{BIRD-Dev} \\
\midrule
\multicolumn{4}{@{}l@{}}{\textit{Exploratory Experiments}} \\
\midrule
Llama3-8B
& \textbf{72.8}
& 62.2
& 32.1 \\
w/ \textit{MCTS}
& 70.1
& \textbf{63.1}
& \textbf{34.2} \\
\midrule
Qwen2.5-7B
& 72.3
& 64.0
& 41.1 \\
w/ \textit{MCTS}
& \textbf{72.5}
& \textbf{65.3}
& \textbf{41.6} \\
\bottomrule
\end{tabularx}
\end{minipage}
\end{table}

\subsection{Ablation Study}
\label{sec:ablation_study}
Each experiment is run three times, and we report the mean with standard error. As shown in Tables~\ref{table:4} and~\ref{table:5}, ablations confirm that SQL-o1's performance depends on the synergy of MCTS, PSG, and cold-start SFT. Removing MCTS significantly reduces accuracy—especially on BIRD-Dev—while excluding SFT causes performance collapse, underscoring the importance of guided initialization. MCTS alone offers limited gains, and Llama3-8B fails without cold-start. Only the full framework ensures stable SQL generation in complex or low-resource scenarios.

\begin{figure}[h]
    \centering
    \begin{minipage}{0.48\textwidth} 
        \centering
        \includegraphics[width=\linewidth]{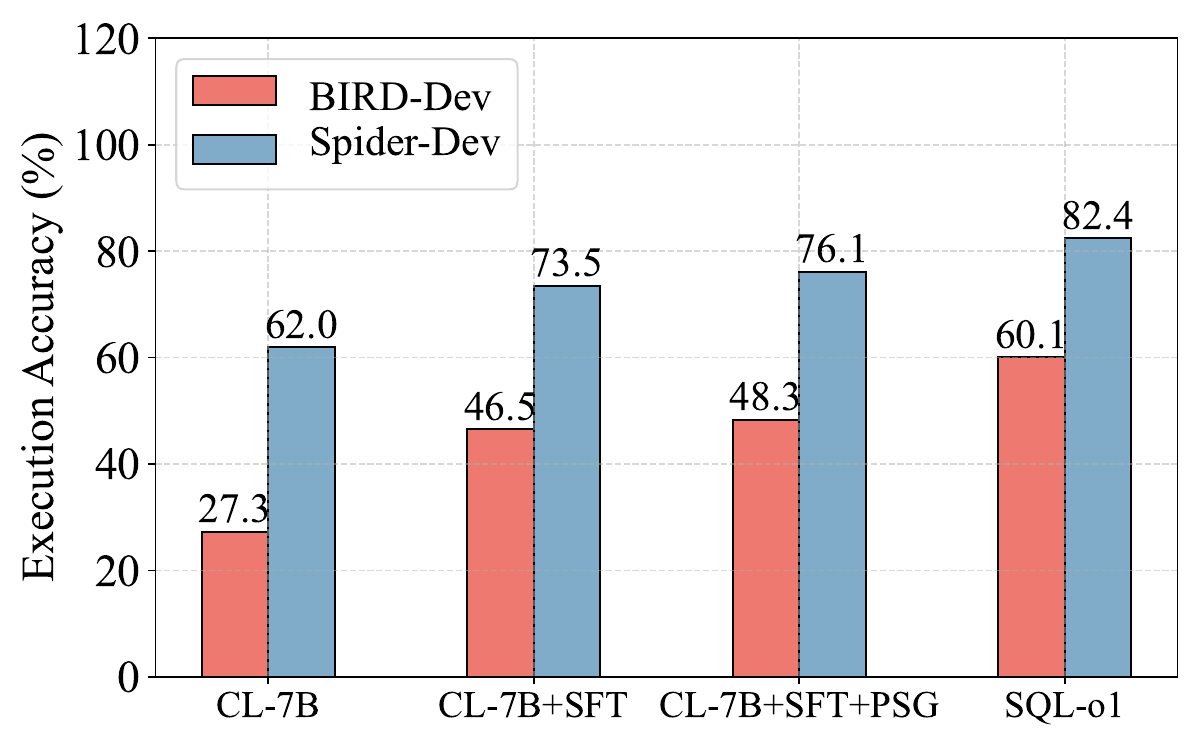}
        \subcaption{CodeLlama-7B (CL-7B).} \label{fig:left}
    \end{minipage}
    \hfill  
    \begin{minipage}{0.48\textwidth}
        \centering
        \includegraphics[width=\linewidth]{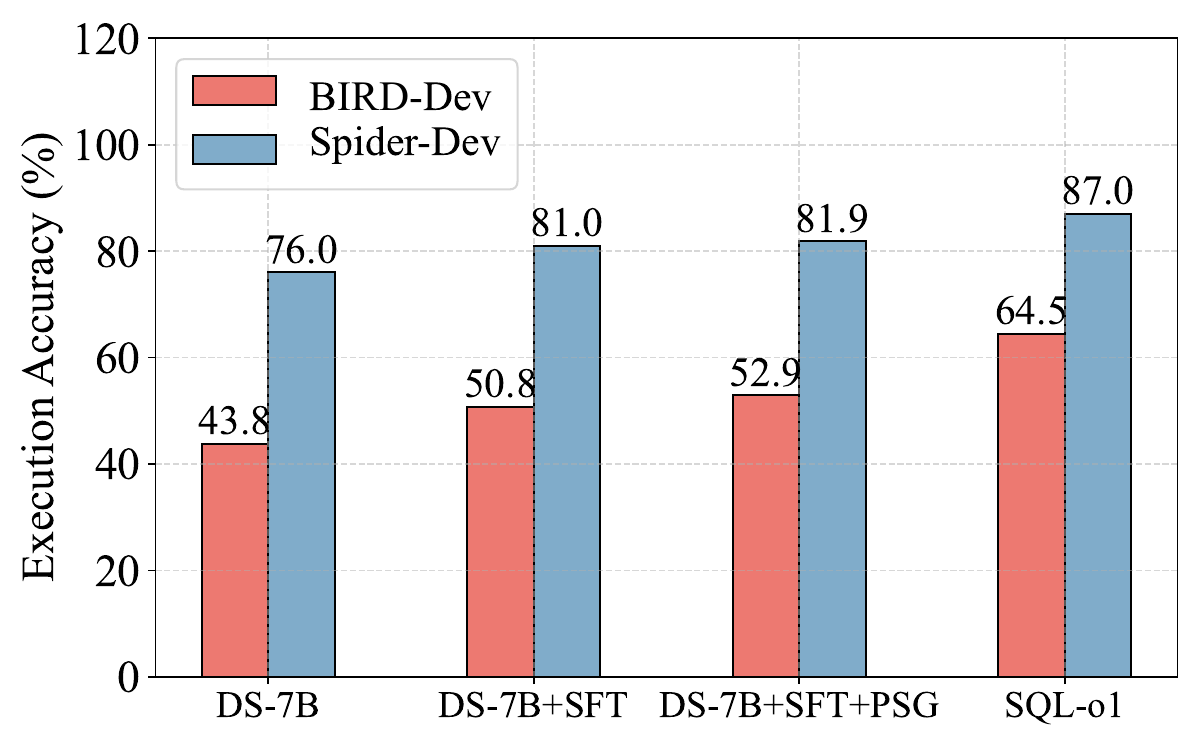}
        \subcaption{DeepSeek-Coder-7B (DS-7B).} \label{fig:right}
    \end{minipage}
    
    \caption{The transferability results on different open-source LLMs on Spider and BIRD.}
    \label{fig:transfer}
    \Description[Comparative bar charts showing model transferability performance]{Two scatter plots placed side by side, labeled (a) and (b), comparing the transferability performance of two large language models, CodeLlama-7B and DeepSeek-Coder-7B, on the Spider and BIRD benchmarks. SQL-o1 consistently achieves state-of-the-art performance.} 
\end{figure}
\vspace{-4mm}

\subsection{Analysis of Transferability}
Figure~\ref{fig:transfer} presents portability results with open-source models such as CodeLlama and DeepSeek-Coder. Without heuristic MCTS, both models suffer performance drops and yield suboptimal results. While trends align with expectations, MCTS gains are larger on BIRD-Dev than Spider-Dev, suggesting generalization of Heuristic Dynamic Search with Self-Reward across model backbones.

\subsection{Logic \& Function Error Reduction with SQL-o1}
\begin{figure}
    \centering
    \includegraphics[width=1.0\linewidth, height=0.5\linewidth]{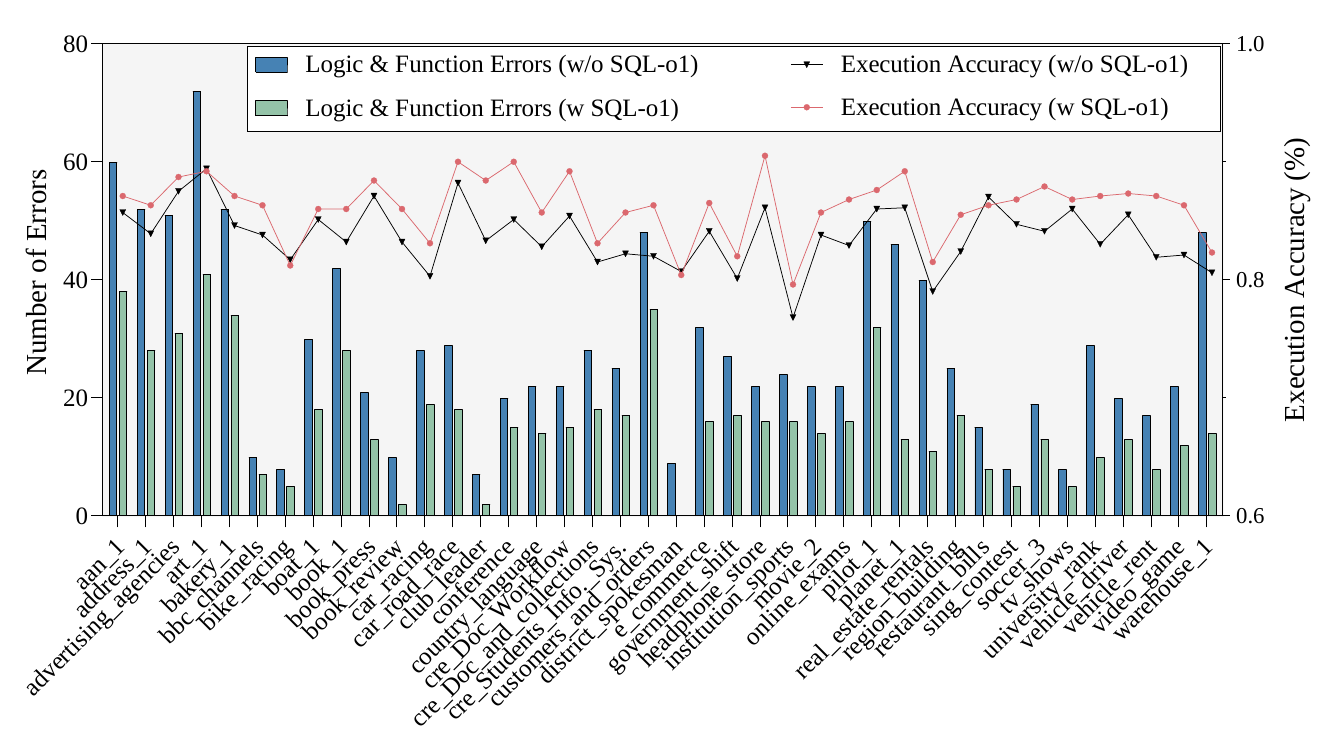}
    \caption{SQL-o1 reduces Logic and Function Errors across databases in the Spider benchmark.}
    \Description[SQL-o1 error reduction]{
    Bar chart showing SQL-o1 significantly reduces both Logic Errors and Function Errors across various database complexity levels in the Spider benchmark. Demonstrates improved accuracy in SQL generation through structured decoding.
    }
    \label{fig:error_reduction}
\end{figure}

We evaluate the effectiveness of SQL-o1 in mitigating Logic \& Function Errors when applied to the open-source Qwen2.5 model on the Spider benchmark. As shown in Figure~\ref{fig:error_reduction}, SQL-o1 leads to a consistent reduction in error counts across different databases. In particular, high-error domains such as \textit{advertising\_agencies}, \textit{bbc}, and \textit{planet} exhibit significant improvements. This highlights the effectiveness of SQL-o1 in enhancing the reasoning capabilities of LLMs, especially in handling complex logic and function usage during SQL generation.

\begin{figure}[bht!]
    \centering
    \begin{minipage}{0.48\columnwidth}  
        \centering
        \includegraphics[width=\linewidth]{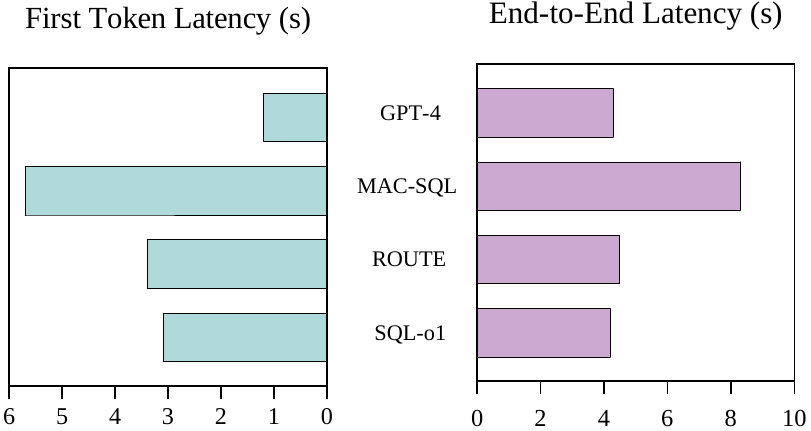}  \linewidth\columnwidth
        \subcaption{Latency Analysis across Methods.}
    \end{minipage}
    \hfill 
    \begin{minipage}{0.46\columnwidth}
        \centering
        \includegraphics[width=0.7\linewidth]{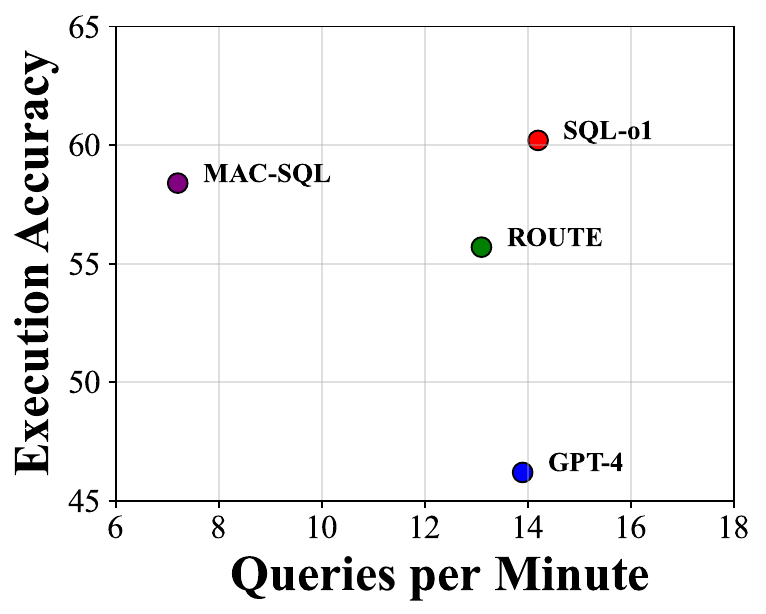}
        \subcaption{Accuracy and Query Efficiency across Methods.}
    \end{minipage}
    \caption{SQL-o1 balances execution accuracy and efficiency across benchmarks.}
    \label{fig:mcts_n_E-A}
\end{figure}
\subsection{Efficiency-Accuracy Trade-offs in Inference Optimization}
\label{sec:Inference_optim}
Inspired by prior work~\cite{ding2025dynamic}, we optimize SQL-o1 to address challenges in inference efficiency. Specifically, by adopting parallel rollout and MCTS dynamic pruning strategies, SQL-o1 effectively expands the generation space while preserving original accuracy, showing clear advantages over baseline methods (Figure~\ref{fig:mcts_n_E-A}). Although ROUTE remains the fastest, SQL-o1 achieves a better efficiency–accuracy trade-off and avoids frequent Decomposer/Refiner calls and their API overhead.

\section{Experimental Details}
\subsection{Dataset Details}
\label{sec:Dataset_detail}
 We use the original training data from Spider and BIRD to construct 16,428 schema-aware pairs. For PSG construction, we perform multi-backbone failure mining using Llama3-8B, Qwen2.5-7B, CodeLlama-7B, and DeepSeek-Coder-7B. Training instances incorrectly predicted by these models are collected into a shared failure pool, with instances exhibiting failures across multiple backbones prioritized. After deduplication, we construct a fixed set of 4,826 syntax-aligned prefix-completion examples. These derived examples are jointly used with the original pairs for cold-start fine-tuning, without introducing additional external SQL annotations.
\subsection{Baseline Details}
\label{sec:baseline_details}
In this section, we summarize all the baseline methods compared in the paper, with a focus on the methods and underlying principles they use. This will help link the evaluation and comparison to the proposed method.

\textbf{DIN-SQL}~\cite{DBLP:conf/nips/PourrezaR23} decomposes the Text-to-SQL task into multiple sub-procedures and applies tailored prompting strategies for each subproblem, guiding GPT-4 to generate the final SQL query through a modular pipeline.

\textbf{DAIL-SQL + GPT-4}~\cite{dail_sql} integrates the lessons learned from comparing question representation and example strategies, selecting the most relevant examples based on the similarity between the question and the query.

\textbf{MAC-SQL}~\cite{wang2024macsql} is based on a multi-agent collaboration framework using LLMs. The MAC-SQL framework consists of a core decomposer agent responsible for generating text-to-SQL, along with two auxiliary agents that utilize external tools or models to retrieve smaller sub-databases and refine incorrect SQL queries.

\textbf{MCS-SQL + GPT-4}~\cite{lee-etal-2025-mcs} utilizes multiple prompts tailored to the database schema to generate candidate SQL queries, and employs a large language model in a multiple-choice setting to select the most confident query based on confidence scores.

\textbf{DTS-SQL + GPT-4}~\cite{DBLP:conf/emnlp/PourrezaR24} proposes a two-stage decomposition framework for text-to-SQL fine-tuning, with a schema-linking pre-generation task before the final SQL generation.

\textbf{CODES}~\cite{DBLP:journals/pacmmod/LiZLFZZWP0024} adopts an incremental pre-training approach with a SQL-centric corpus, tackling schema linking and rapid domain adaptation challenges through strategic prompt construction and bidirectional data augmentation techniques.

\textbf{SENSE}~\cite{yang-etal-2024-synthesizing} introduces a synthetic data method that combines strong and weak data, using preference learning to help the model learn from mistakes, thereby improving the performance of open-source LLMs on the text-to-SQL task.

\textbf{ROUTE}~\cite{qin2024route} introduces multiple fine-tuning and task collaboration prompt (MCP) strategies. These strategies leverage collaboration between SQL-related tasks to reduce hallucinations during the SQL generation process.

\textbf{C3-SQL + ChatGPT}~\cite{DBLP:journals/corr/abs-2307-07306} adopts ChatGPT’s zero-shot text-to-SQL method, C3, with approximately 1000 tokens per query. The framework comprises three key components: Clear Prompt (CP), Prompt Calibration (CH), and Consistent Output (CO), where CH serves to reduce potential bias and enhance performance.

\textbf{RESDSQL}~\cite{DBLP:conf/aaai/Li00023} is a framework for the text-to-SQL task, designed to improve performance and robustness by decoupling schema linking and skeleton parsing.

\textbf{CHASE-SQL}~\cite{pourreza2024chase} introduces query decomposition, execution-aware reasoning, instance-specific examples, and candidate ranking to improve the diversity and accuracy of SQL generation.

\textbf{XiYan-SQL}~\cite{gao2024xiyan} integrates a multi-generator ensemble strategy with the M-Schema method for database representation, and adopts a three-stage process comprising schema linking, SQL generation and optimization, and SQL selection. It demonstrates superior performance on benchmark datasets such as Spider and SQL-Eval.

\textbf{CHESS} employs more sophisticated extraction patterns to identify key fields, and excels in handling complex SQL queries.
\begin{figure}[t!]
\centering
\includegraphics[width=0.85\linewidth]{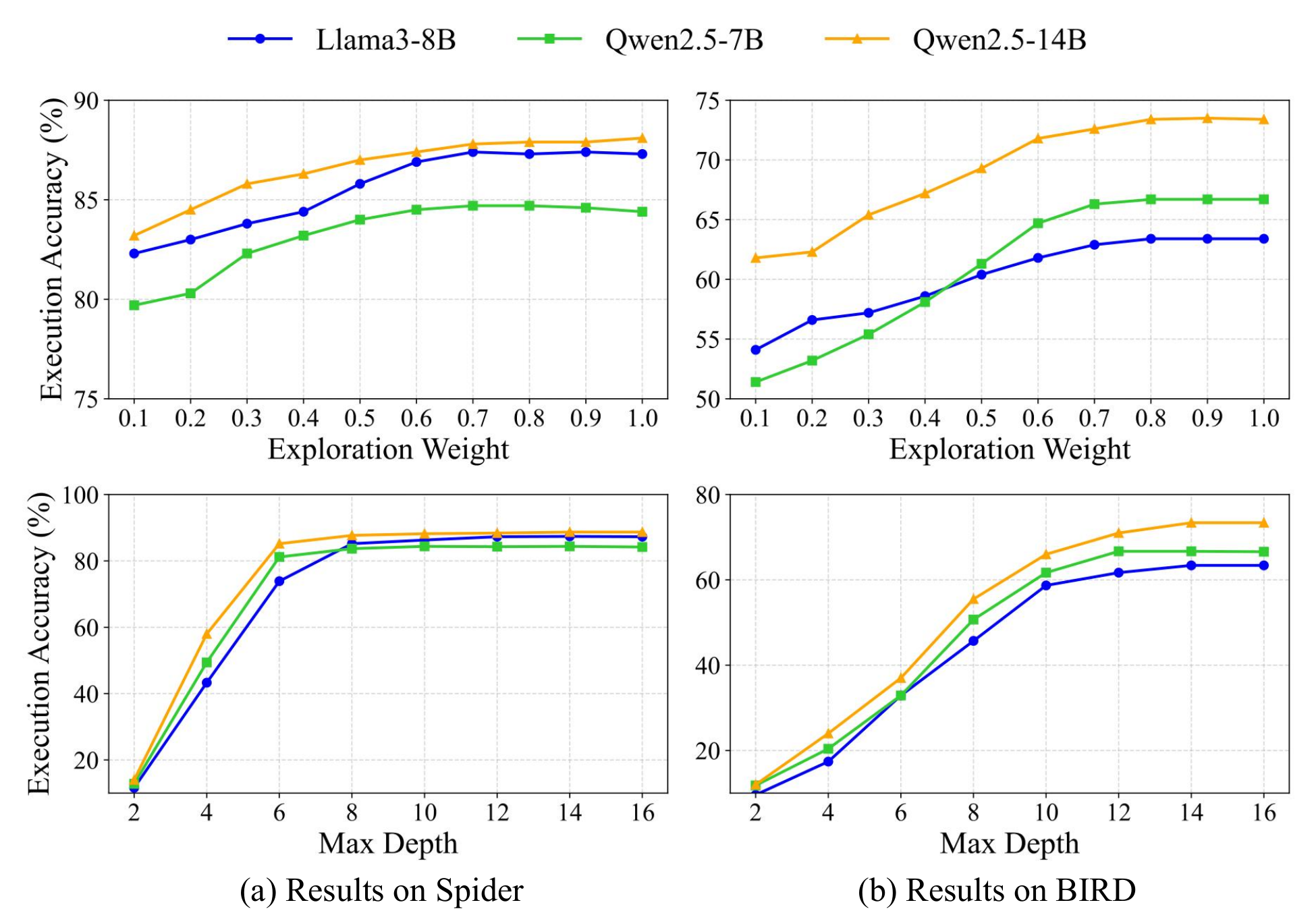}  
\caption{Analysis of exploration weight and search depth effects on Spider and BIRD datasets.}
\label{fig:hyperparameter}
\end{figure}
\vspace{-2mm}
\subsection{Implementation Details}

\paragraph{Experimental Setup.}
Experiments are run on 8 NVIDIA A40 GPUs (48GB) with a batch size of 32 using the Llama-Factory framework. We evaluate open-source LLMs including Llama3 and Qwen2.5. The learning rate starts at \(1 \times 10^{-5}\) and decays to zero via a cosine schedule. MCTS rollout counts are set to 6 for Spider and 8 for BIRD. The reward temperature $\alpha$ is set to 0.6. All models are fine-tuned for 2 epochs with gradient clipping at 1.0 to stabilize training. Ablation results are averaged over three runs with different seeds to ensure statistical reliability and reduce variance due to random initialization.

\paragraph{Inference Efficiency Evaluation.} We evaluate inference efficiency on fixed random subsets of 200 examples from Spider-Dev and BIRD-Dev, respectively, consistent with the evaluation scale used in our threshold-sensitivity analysis. Open-source models are evaluated on a machine equipped with 8 NVIDIA A40 GPUs (48GB each), using an inference batch size of 1 to measure per-query latency. Each configuration is evaluated over three repeated runs, and the average first-token latency, end-to-end latency, and query throughput are reported. For API-based methods, latency is measured from request submission to receipt of the complete SQL output, including network transmission and remote-service overhead. Within each dataset, all methods are evaluated using the same query subset and measurement protocol.

\begin{figure}
    \centering
    \begin{minipage}{0.48\columnwidth}
        \centering
        \includegraphics[width=\linewidth]{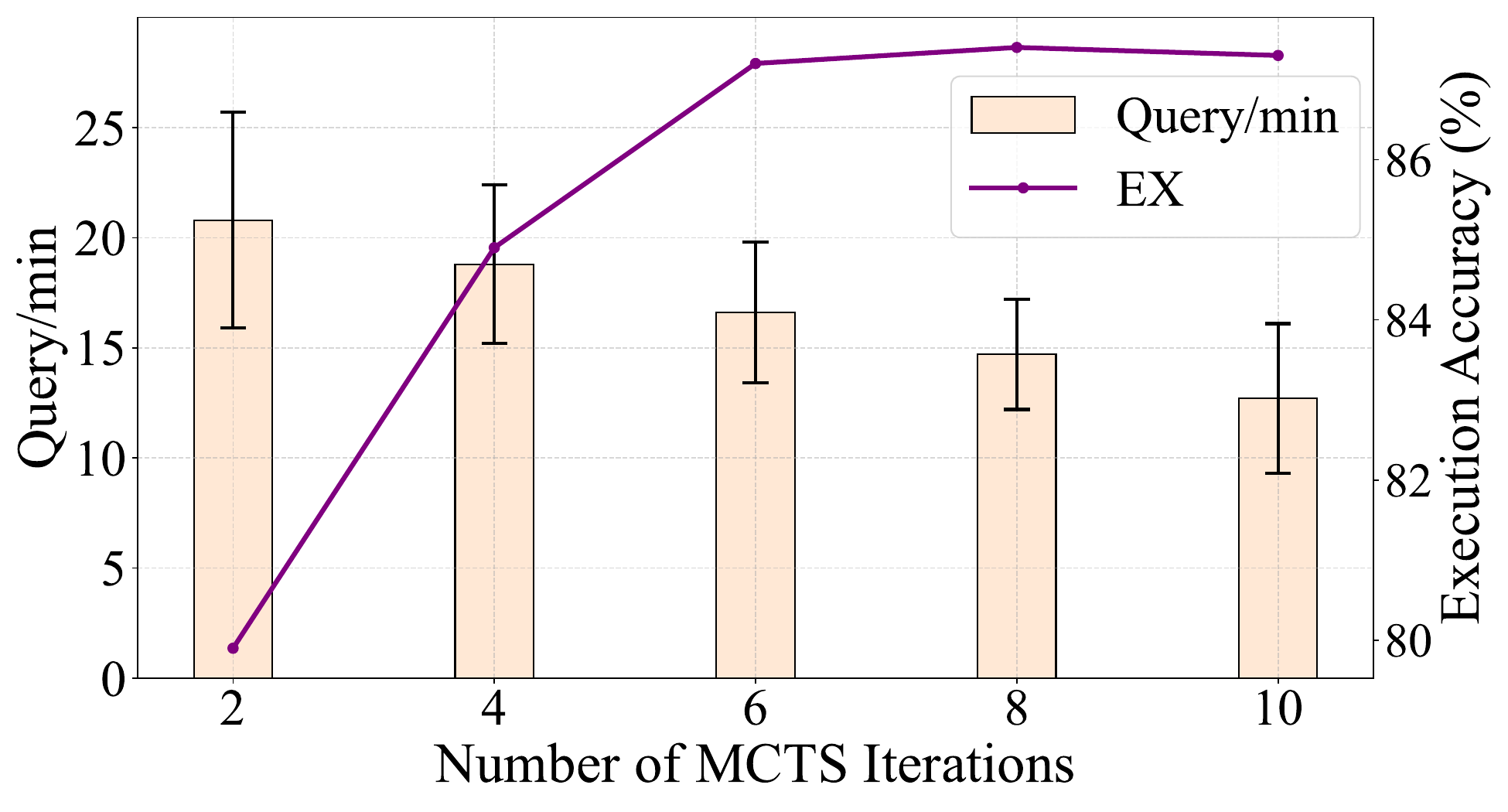}
        \subcaption{Spider-Dev} 
        \label{fig:subfig1} 
    \end{minipage}
    \hfill
    \begin{minipage}{0.48\columnwidth}
        \centering
        \includegraphics[width=\linewidth]{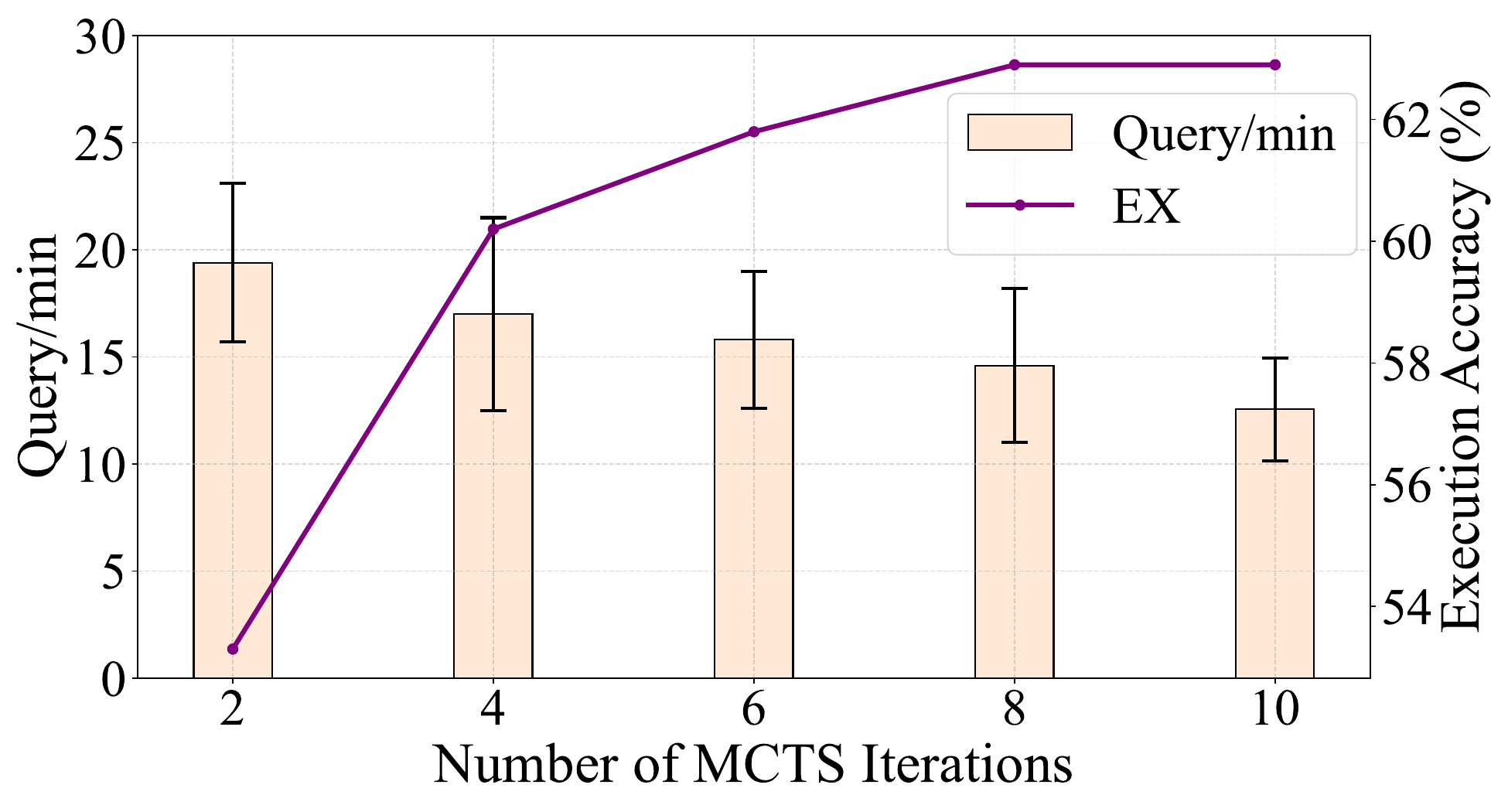}
        \subcaption{BIRD-Dev} 
        \label{fig:subfig2} 
    \end{minipage}
    
    \caption{Effect of MCTS iterations on execution accuracy and inference efficiency.}
    \label{fig:mcts_n}
    \Description[Two side-by-side charts showing MCTS iteration effects]{Left chart shows iteration effects on Spider-Dev dataset, right chart shows effects on BIRD-Dev dataset, both measuring execution accuracy and inference efficiency across different iteration counts.} 
\end{figure}

\paragraph{Hyperparameter Settings.}
Beyond the \textit{Rollout Count} analysis in Figure~\ref{fig:mcts_n}, we further examine the impact of \textit{Exploration Weight} and \textit{Maximum Depth}, with results shown in Figure~\ref{fig:hyperparameter}. Table~\ref{tab:hyperparameter} lists the hyperparameter configurations for Spider and BIRD, divided into two stages: Initial Fine-tuning and MCTS Exploration, both tailored to optimize performance on each dataset.
\begin{table}
\centering
\caption{Hyperparameter Settings for Spider and BIRD datasets.}
\label{tab:hyperparameter}
\footnotesize
\setlength{\tabcolsep}{4pt}
\renewcommand{\arraystretch}{0.92}
\setlength{\belowcaptionskip}{3pt}
\begin{tabular*}{\linewidth}{l@{\extracolsep{\fill}}cc}
\toprule
\textbf{Hyperparameter Name} & \textbf{Spider} & \textbf{BIRD} \\
\midrule
\multicolumn{3}{c}{\textit{\textbf{Initial Fine-tuning Stage}}} \\
\midrule
SFT Type & LoRA & LoRA \\
Batch Size & 32 & 32 \\
Learning Rate & 1e-5 & 1e-5 \\
Training Epochs & 2 & 2 \\
Progressive SQL Generation Data Ratio & 22.7\% & 22.7\% \\
\midrule
\multicolumn{3}{c}{\textit{\textbf{MCTS Exploration Stage}}} \\
\midrule
Rollout Count ($N$) & 6 & 8 \\
Beam Size ($B$) & 5 & 5 \\
Action Space Size ($k$) & 3 & 3 \\
Maximum Depth ($L$) & 8 & 12 \\
Exploration Weight ($w$) & 0.7 & 0.8 \\
Reward Temperature ($\alpha$) & 0.6 & 0.6 \\
Base Reward Score ($\beta$) & 100 & 100 \\
Reward Balance Ratio ($\delta$) & 0.5 & 0.5 \\
Similarity Threshold & 0.7 & 0.7 \\
\bottomrule
\end{tabular*}
\end{table}

In the Initial Fine-tuning Stage, both the Spider and BIRD datasets use LoRA-based training. The batch size is set to 32, the learning rate to 1e-5, and training is conducted for 2 epochs. The PSG prefix-completion examples account for 22.7\% of the cold-start training data.

In the MCTS Exploration Stage, dataset-specific search configurations are used for Spider and BIRD. We set the Rollout Count (N) to 6 for the Spider dataset and 8 for the BIRD dataset, with each exploration stage simulating multiple paths to collect potential solutions. The Beam Size (B) is set to 5, limiting the number of candidate paths to the top 5. The Action Space Size (k) is set to 3, indicating the number of possible actions considered at each decision point. The Maximum Depth (L) is set to 8 for Spider and 12 for BIRD, reflecting the depth of decision tree exploration.

The Exploration Weight ($w$) is set to 0.7 for Spider and 0.8 for BIRD, respectively. The Reward Temperature ($\alpha$) is set to 0.6, influencing the model’s sensitivity to reward signals. The Base Reward Score ($\beta$) is set to 100 for both datasets, while the Reward Balance Ratio ($\delta$) is set to 0.5 to balance the process-level and terminal reward signals during MCTS search. Finally, the similarity threshold used to filter redundant or irrelevant paths is defined as 0.7 for both datasets.

\begin{figure}
    \centering
    \includegraphics[width=0.9\linewidth]{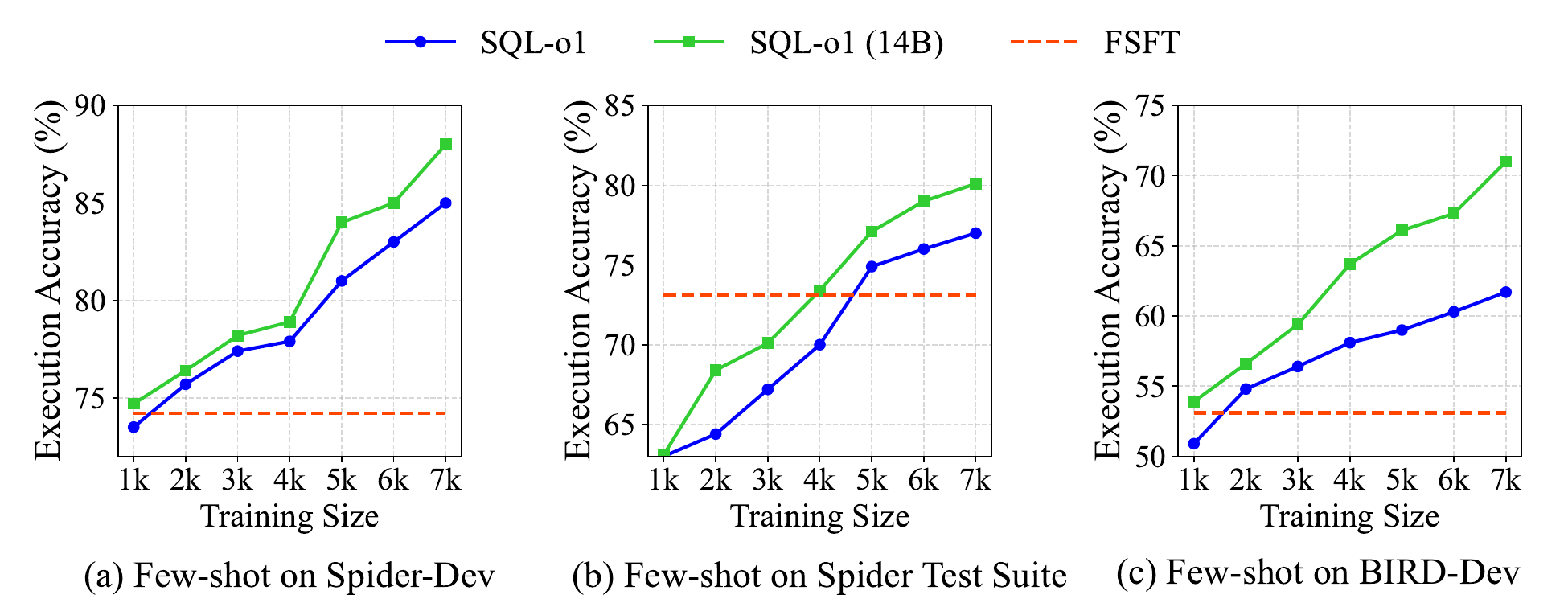}
    \caption{Performance comparison of SQL-o1 and Fully Supervised Fine-Tuned (FSFT) Llama3-8B on different datasets and sample sizes.}
    \label{fig:fewsample}
\end{figure}

\paragraph{Model Performance in Low-Resource Tasks}
We compare SQL-o1 with a fully fine-tuned Llama3-8B on the Spider and BIRD datasets (Spider-Dev-EX, Spider-Dev-TS, BIRD-Dev-EX). As shown in Figure~\ref{fig:fewsample}, SQL-o1 outperforms Llama3-8B when trained on small sample sizes (2000–5000 examples), and its advantage becomes more pronounced on Spider-Dev-EX and BIRD-Dev as the number of training samples increases. SQL-o1 excels in few-shot learning by leveraging limited supervision to generalize across complex query structures, demonstrating its superiority in resource-constrained scenarios and reasoning capabilities under limited data conditions.

\subsection{Algorithm Details}
During \textit{Selection}, nodes are chosen via the Upper Confidence Bound for Trees (UCT), balancing exploration and exploitation. In \textit{Expansion}, the policy network $\pi_\theta$ generates candidate SQL fragments conditioned on the current state, filtered by schema-aware beam search before being added as child nodes. The Simulation phase completes these into full SQL queries and computes the rollout return $G^{(n)}$ from the cumulative process reward and execution feedback. In Back-propagation, $G^{(n)}$ is propagated upward to update node values $V(u)$ and visit counts $N(u)$. The algorithmic details are provided in Algorithm~\ref{alg:mcts}.

\begin{algorithm}[htb!]
\caption{MCTS in SQL-o1}
\label{alg:mcts}
\footnotesize
\begin{algorithmic}[1]
\REQUIRE Initial state $s_0$, schema-aware policy model $\pi_{\theta}$ (serving as both policy and reward model), beam size $B$, number of rollouts $N$, maximum search depth $L$, exploration weight $w$, reward balance ratio $\delta$, pruning threshold schedule $\{\tau_t\}_{t=1}^{L}$
\ENSURE A fully generated SQL query $S$
\STATE Initialize search tree $\mathcal{T}$ with root node $v_0$ corresponding to initial state $s_0$.
\STATE Initialize the set of previously observed non-empty normalized execution results $\mathcal{Z}_0 \leftarrow \emptyset$.
\FOR{$n \leftarrow 1 \text{ to } N$}
    \STATE \textbf{Selection:}
    \STATE \quad Set current node $v \leftarrow v_0$.
    \STATE \quad \textbf{while} $v$ is not a leaf node and search depth $< L$ \textbf{do}
    \STATE \quad\quad $v \leftarrow \arg\max_{v' \in \text{children}(v)} \left[ V(v') + w \cdot \sqrt{\frac{\ln N(v)}{N(v')}} \right]$
    \STATE \quad \textbf{end while}
    \STATE \textbf{Expansion:}
    \STATE \quad \textbf{if} $v$ is not a terminal node \textbf{then}
    \STATE \quad\quad Use $\pi_{\theta}$ to generate up to $B$ candidate actions $\{e^b\}_{b=1}^{B}$ from the current state.
    \STATE \quad\quad Compute model confidence scores $\{s_b\}_{b=1}^{B}$ for each $e^b$.
    \STATE \quad\quad \textit{// $s_b$ denotes the model-assigned confidence score for candidate action $e^b$}
    \STATE \quad\quad \textit{// $\tau_t$ is the dynamic pruning threshold at step $t$ (see Equation~(10))}
    \STATE \quad\quad \textbf{Dynamic Pruning:} Retain only candidates where $s_b \ge \tau_t$.
    \STATE \quad\quad Keep top $d \le B$ actions via beam filtering, create new child nodes $\{v^b\}_{b=1}^{d}$ for each action.
    \STATE \quad \textbf{end if}
    \STATE \textbf{Simulation:}
    \STATE \quad Select a child node $v^*$ from $\{v^b\}_{b=1}^{d}$ using a heuristic (e.g., highest beam score).
    \STATE \quad Simulate generation with $\pi_{\theta}$ until reaching depth $L$ or a terminal SQL query $S^{(n)}$.
    \STATE \quad Compute the step-level self-rewards $r_t^{(n)}$ along the simulated trajectory and aggregate them into the cumulative process reward $R_{\mathrm{cum}}^{(n)}$.
    \STATE \quad If a terminal SQL query $S^{(n)}$ is generated, execute it on the target database and compute $R_{\mathrm{exec}}(S^{(n)})$.
    \STATE \quad Set $R_{\mathrm{exec}}(S^{(n)})=-1$ if SQL execution fails.
    \STATE \quad Otherwise, let $z^{(n)}=\mathrm{NormExec}(S^{(n)},\mathcal{D})$ and set $R_{\mathrm{exec}}(S^{(n)})=+1$ if $z^{(n)}$ is non-empty and $z^{(n)}\in\mathcal{Z}_{n-1}$, and $0$ otherwise.
    \STATE \quad If $z^{(n)}$ is non-empty, update $\mathcal{Z}_{n}\leftarrow\mathcal{Z}_{n-1}\cup\{z^{(n)}\}$; otherwise set $\mathcal{Z}_{n}\leftarrow\mathcal{Z}_{n-1}$.
    \STATE \quad Compute the rollout return $G^{(n)}$ according to Equation~(7).
    \STATE \textbf{Back-propagation:}
    \STATE \quad Let $\tau = (v_0 \rightarrow \dots \rightarrow v^*)$ denote the selected trajectory from the root to the simulated leaf node.
    \STATE \quad For each node $u$ in $\tau$:
    \STATE \quad\quad $N(u) \leftarrow N(u) + 1$
    \STATE \quad\quad $V(u) \leftarrow V(u) + \frac{G^{(n)}-V(u)}{N(u)}$
\ENDFOR
\STATE \textbf{Return} the SQL query from the child of $v_0$ with the highest $V(\cdot)$ value.
\end{algorithmic}
\end{algorithm}

\subsubsection{Complexity Analysis of SQL-o1}
\label{sec:Complexity_Ana}
The time complexity of the MCTS process in SQL-o1 can be analyzed through its four phases: selection, expansion, simulation, and Back-propagation. In the selection phase, the algorithm traverses the search tree up to depth $L$ and selects a node using the UCT criterion. The time complexity per rollout is $O(k \cdot L)$, where $k$ is the number of possible actions at each step. During expansion, the model generates $B$ candidate beams and filters them based on similarity, introducing an additional complexity of $O(k \cdot \omega)$, where $\omega \in [0, 1]$. The simulation phase explores the path to depth $L$, contributing a complexity of $O(k \cdot L)$. In Back-propagation, rewards are updated along the same path, requiring $O(L)$ time. These estimates reflect the worst-case complexity; in practice, dynamic pruning may reduce computation by terminating unpromising branches before the maximum search depth is reached.

Each rollout has complexity $O(kL)$; hence, the total complexity of MCTS for $N$ rollouts is $O(NkL)$, scaling linearly with the number of rollouts and search depth.

\subsubsection{Improving MCTS Efficiency in SQL-o1}

\begin{table}
\centering
\caption{Impact of the hyperparameter $\lambda$ on inference efficiency}
\label{table:lambda_impact}
\footnotesize
\setlength{\tabcolsep}{5pt}
\renewcommand{\arraystretch}{0.95}
\setlength{\belowcaptionskip}{3pt}
\begin{tabular}{cccc}
\toprule
\textbf{$\lambda$} & \textbf{EX} & \textbf{Avg. Time (s)} & \textbf{Pruned Paths (\%)} \\
\midrule
1.0  & 78.5 & 4.18  & 42.1 \\
0.95 & 80.7 & 4.21  & 27.7 \\
0.9  & 82.6 & 4.32  & 20.8 \\
0.85 & 82.4 & 5.24  & 17.1 \\
0.8  & 81.8 & 8.21  & 14.6 \\
0.6  & 80.3 & 7.38  & 7.3  \\
0.4  & 81.6 & 4.21  & 0.2  \\
0.2  & 81.0 & 8.45  & 0    \\
0    & 82.3 & 10.78 & 0    \\
\bottomrule
\end{tabular}
\end{table}

We randomly sample 200 examples from the validation sets of Spider and BIRD to evaluate the impact of the threshold hyperparameter~$\lambda$ on inference efficiency and pruning behavior. As shown in Table~\ref{table:lambda_impact}, higher values of $\lambda$ lead to more aggressive pruning, significantly reducing average inference time. For example, when $\lambda = 1.0$, over 42\% of candidate paths are pruned early, leading to the fastest inference but at the cost of lower execution accuracy (EX). In contrast, smaller $\lambda$ values reduce the aggressiveness of pruning, enabling deeper exploration but also introducing higher latency. When $\lambda = 0.9$, the model achieves the best trade-off between accuracy and efficiency. These results suggest that tuning $\lambda$ allows flexible control over the accuracy-efficiency balance, making it possible to adapt SQL-o1 to different latency constraints or deployment scenarios.

\subsubsection{Effect of MCTS Rollout Count on Inference Efficiency}
\label{sec:MCTS_iter}
Figure~\ref{fig:mcts_n} illustrates the impact of the optimized number of MCTS iterations on the number of queries processed per minute during heuristic search. We evaluate SQL-o1 with Llama3-8B on the Spider-Dev and BIRD-Dev datasets. The results show that the EX metric consistently improves as the number of iterations increases, reflecting the fundamental characteristic of heuristic search: the accuracy of SQL generation is bounded by the search space, which expands proportionally with the number of search attempts. On the Spider dataset, the EX metric peaks at 6 iterations and achieves approximately 16.3 queries per minute. 
Further iterations yield diminishing returns. On the BIRD dataset, the EX metric at 8 iterations approaches its peak value.

\begin{table}[htb!]
\centering
\caption{Performance (EX) comparison under different hardness levels on the Spider development set. \textbf{Bold} numbers represent the highest performance among fine-tuning-based methods.}
\label{table:A_1}
\footnotesize
\setlength{\tabcolsep}{4pt}
\renewcommand{\arraystretch}{1.0}
\begin{tabular}{lccccc}
\hline
\textbf{Methods} & \textbf{Easy} & \textbf{Medium} & \textbf{Hard} & \textbf{Extra} & \textbf{All} \\
\hline
\multicolumn{6}{c}{\textit{\textbf{Prompting-based methods}}} \\
\hline
C3-SQL + ChatGPT~\cite{DBLP:journals/corr/abs-2307-07306} & 92.7 & 85.2 & 77.6 & 62.0 & 82.0 \\
DAIL-SQL + GPT-4 + Self-consistency & 91.5 & 90.1 & 75.3 & 62.7 & 83.6 \\
SuperSQL + GPT-4~\cite{Li_2024} & 94.4 & 91.3 & 83.3 & 68.7 & 87.0 \\
MCS-SQL + GPT-4~\cite{lee-etal-2025-mcs} & 94.0 & 93.5 & 88.5 & 72.9 & 89.5 \\
\hline
\multicolumn{6}{c}{\textit{\textbf{Fine-tuning-based methods}}} \\
\hline
DTS-SQL~\cite{DBLP:conf/emnlp/PourrezaR24} & 92.7 & 90.1 & 74.1 & 56.6 & 82.7 \\
RESDSQL-3B + NatSQL~\cite{DBLP:conf/aaai/Li00023} & 94.4 & 87.9 & 77.0 & 66.3 & 84.1 \\
\rowcolor{gray!15}
ROUTE + Llama3-8B~\cite{qin2024route} & \textbf{96.0} & \textbf{93.0} & 75.3 & 63.3 & 86.0 \\
\rowcolor{gray!15}
ROUTE + Qwen2.5-7B~\cite{qin2024route} & 92.7 & 89.7 & 77.0 & 60.2 & 83.6 \\
\rowcolor{gray!15}
ROUTE + Qwen2.5-14B~\cite{qin2024route} & 94.0 & \textbf{93.0} & \textbf{81.6} & 68.1 & 87.3 \\
\rowcolor{cyan!15}
Ours: SQL-o1 + Llama3-8B & 94.4 & \textbf{93.0} & 81.0 & \textbf{68.7} & \textbf{87.4} \\
\hline
\end{tabular}
\end{table}

\begin{table}[htb!]
\centering
\caption{Performance (EX) comparison under different hardness levels on the BIRD development set. \textbf{Bold} numbers represent the highest performance among fine-tuning-based methods.}
\label{table:A_2}
\footnotesize
\setlength{\tabcolsep}{4pt}
\renewcommand{\arraystretch}{1.0}
\begin{tabular}{lcccc}
\hline
\textbf{Methods} & \textbf{Simple} & \textbf{Moderate} & \textbf{Challenging} & \textbf{All} \\
\hline
\multicolumn{5}{c}{\textit{\textbf{Prompting-based methods}}} \\
\hline
C3-SQL + ChatGPT~\cite{DBLP:journals/corr/abs-2307-07306} & 58.9 & 38.5 & 31.9 & 50.2 \\
DAIL-SQL + GPT-4 & 62.5 & 43.2 & 37.5 & 54.3 \\
DAIL-SQL + GPT-4 + Self-consistency & 63.0 & 45.6 & 43.1 & 55.9 \\
SuperSQL + GPT-4~\cite{Li_2024} & 66.9 & 46.5 & 43.8 & 58.5 \\
MAC-SQL + GPT-4~\cite{wang2024macsql} & 65.7 & 52.7 & 40.3 & 59.4 \\
MCS-SQL + GPT-4~\cite{lee-etal-2025-mcs} & 70.4 & 53.1 & 51.4 & 63.4 \\
\hline
\multicolumn{5}{c}{\textit{\textbf{Fine-tuning-based methods}}} \\
\hline
RESDSQL-3B~\cite{DBLP:conf/aaai/Li00023} & 53.5 & 33.3 & 16.7 & 43.9 \\
CodES-7B + SFT~\cite{DBLP:journals/pacmmod/LiZLFZZWP0024} & 64.6 & 46.9 & 40.3 & 57.0 \\
CodES-15B + SFT~\cite{DBLP:journals/pacmmod/LiZLFZZWP0024} & 65.8 & 48.8 & 42.4 & 58.5 \\
ROUTE + Llama3-8B~\cite{qin2024route} & 64.3 & 49.3 & 36.8 & 57.3 \\
\rowcolor{gray!15}
ROUTE + Qwen2.5-7B~\cite{qin2024route} & 63.8 & 45.4 & 39.6 & 55.9 \\
\rowcolor{gray!15}
ROUTE + Qwen2.5-14B~\cite{qin2024route} & 67.7 & \textbf{53.1} & 42.4 & 60.9 \\
\rowcolor{cyan!15}
Ours: SQL-o1 + Llama3-8B & \textbf{71.8} & 52.3 & \textbf{45.2} & \textbf{63.4} \\
\hline
\end{tabular}
\end{table}

\subsection{Comparison of Performance with Different Hardness}
To enable detailed comparison, we report SQL-o1’s EX scores on the Spider and BIRD dev sets across query difficulty levels. As shown in Tables~\ref{table:A_1} and~\ref{table:A_2}, SQL-o1 consistently ranks among the top models at all levels, with fine-grained metrics confirming its robustness and stability.

\subsection{Scaling SQL-o1 to Larger Open-Source Language Models}
\label{sec:appendix:compare_gpt}
To examine scaling under a consistent backbone family, Table~\ref{table:G_1} compares ROUTE, ROUTE+MCP, and SQL-o1 with both Qwen2.5-7B and Qwen2.5-14B. SQL-o1 improves from 84.7 to 88.4 on Spider-Dev and from 66.7 to 73.1 on BIRD-Dev when scaling from 7B to 14B, while maintaining its advantage over the corresponding same-size baselines.

Table~\ref{table:G_1} also includes representative proprietary systems for reference. SQL-o1 with Qwen2.5-14B achieves 88.4 Spider-Dev-EX and 73.1 BIRD-Dev-EX, which are comparable to CHASE-SQL and XiYan-SQL.

\begin{table}[t]
\centering
\caption{Scaling comparison of SQL-o1 with same-backbone open-source baselines and representative proprietary Text-to-SQL systems. \textbf{Bold} numbers represent the highest performance.}
\label{table:G_1}
\footnotesize
\setlength{\tabcolsep}{4pt}
\renewcommand{\arraystretch}{1.08}

\begin{tabularx}{\linewidth}{
@{}
>{\raggedright\arraybackslash}p{2.0cm}
>{\raggedright\arraybackslash}p{2.0cm}
*{3}{>{\centering\arraybackslash}X}
@{}
}
\toprule
\textbf{Methods} & \textbf{LLMs} & \textbf{SPIDER-Dev-EX} & \textbf{BIRD-Dev-EX} & \textbf{BIRD-Dev-VES} \\
\midrule
\multicolumn{5}{c}{\textit{\textbf{Proprietary reference systems}}} \\
\midrule
CHASE-SQL~\cite{pourreza2024chase} & Gemini 1.5 Pro & 87.6 & 73.1 & 73.0 \\
XiYan-SQL~\cite{gao2024xiyan} & GPT-4 & -- & \textbf{73.3} & -- \\
\midrule
\multicolumn{5}{c}{\textit{\textbf{Same-backbone open-source comparisons}}} \\
\midrule
ROUTE+MCP & Qwen2.5-7B & 78.3 & 49.7 & 52.8 \\
ROUTE+MCP & Qwen2.5-14B & 80.0 & 56.3 & 57.6 \\
\rowcolor{gray!15}
ROUTE & Qwen2.5-7B & 83.6 & 55.9 & 57.4 \\
\rowcolor{gray!15}
ROUTE & Qwen2.5-14B & 87.3 & 60.9 & -- \\
\rowcolor{cyan!15}
Ours: SQL-o1 & Qwen2.5-7B & 84.7 & 66.7 & 70.4 \\
\rowcolor{cyan!15}
Ours: SQL-o1 & Qwen2.5-14B & \textbf{88.4} & 73.1 & \textbf{75.8} \\
\bottomrule
\end{tabularx}
\end{table}

\subsection{Token Consumption and Cost}

To facilitate a more rigorous and transparent cost comparison, we extend our economic analysis by evaluating inference expenditures under a standardized open-source deployment scenario. In response to reviewer suggestions, we adopt TogetherAI as a representative cloud platform for hosting open-weight models. Our experimental setup provisions eight NVIDIA H100 GPUs, incurring an average rental cost of \$2.85 per hour — a configuration selected to ensure consistent throughput and reproducibility.

\begin{table}[hbt!]
\centering
\caption{Comparison of model performance and inference cost on the Spider development set.}
\label{table_cost}
\resizebox{\textwidth}{!}{
\begin{tabular}{llcccc}
\hline
\textbf{Methods} & \textbf{Base Model} & 
\begin{tabular}{c}\textbf{Average Token} \\ \textbf{Consumption}\end{tabular} &
\begin{tabular}{c}\textbf{Total Cost} \\ \textbf{(\$)}\end{tabular} &
\begin{tabular}{c}\textbf{Cost per} \\ \textbf{1K Tokens (\$)}\end{tabular} &
\begin{tabular}{c}\textbf{Spider-Dev} \\ \textbf{(EX)}\end{tabular} \\
\hline
XiYan-SQL & Multiple models & 6,000 & 156.91 & 0.02529 & 89.6 \\
CSC-SQL ($n=64$) & Qwen2.5-Coder-7B & 9,600 & 6.16 & 0.00062 & 69.2 \\
ROUTE & Qwen2.5-14B & 14,531 & 3.28 & 0.00022 & 87.3 \\
\rowcolor{cyan!15}
Ours: SQL-o1 & Qwen2.5-7B & 9,683 & 3.24 & 0.00032 & 84.7 \\
\rowcolor{cyan!15}
Ours: SQL-o1 & Qwen2.5-14B & 15,768 & 3.48 & 0.00021 & 88.4 \\
\hline
\end{tabular}}
\Description[Comparison of model performance and inference cost.]{
This table compares XiYan-SQL, CSC-SQL, ROUTE, and SQL-o1 in terms of
base model, average token consumption, total inference cost, average cost
per token, and execution accuracy on the Spider development set.
}
\end{table}

Under this regime, we recompute the token-level inference costs for SQL-o1 using two variants of the Qwen2.5 series: the 14B and 7B parameter models. The resulting normalized costs are approximately \$0.00021 and \$0.00032 per 1K tokens for the 14B and 7B models, respectively. These figures reflect substantial efficiency gains compared to systems reliant on commercial APIs. For context, XiYan-SQL — which integrates GPT-4o alongside large open-source coders (e.g., Qwen2.5-Coder-32B/7B) — incurs a total cost of \$156.91 (Table \ref{table_cost}) over the same evaluation set, with \$153.60 (97.9\%) attributable to proprietary API calls.

To promote cross-method comparability and architectural transparency, we retain the original cost table (based on closed-source API pricing) while introducing a normalized metric: \textit{Cost per 1K Tokens (\$)}. This measure enables direct economic benchmarking across heterogeneous model families and deployment strategies, supporting informed design trade-offs in both research and production settings.

\begin{figure}[h] 
  \centering
  \includegraphics[width=0.4\textwidth]{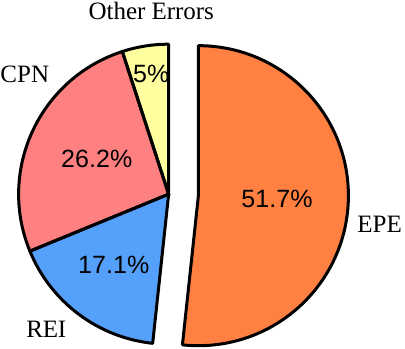} 
  \caption{Proportions of SQL-o1 error types on BIRD (EPE: Execution-Path Errors; REI: Reward Evaluation Inconsistency; CPN: Correct Path Not Discovered; OE: Other Errors).}
  \label{fig:Error_Types}
  \Description[A bar chart showing the distribution of different error types.]{A detailed bar chart titled "Proportions of SQL-o1 Error Types". It shows four categories: Execution Errors (EPE), Reward Inconsistency (REI), Missed Correct Path (CPN), and Other (OE), with EPE being the most common.}
  \vspace{-2mm}
\end{figure}

\subsection{Error Analysis}
We analyze the errors made by SQL-o1 on the BIRD dataset, as illustrated in Figure~\ref{fig:Error_Types}. These errors are categorized into four main types.

\textbf{Execution errors in the exploration path (51.7\%)} These occur when candidate paths generated during MCTS fail due to syntax errors or missing conditions such as filtering, joining, or aggregation, resulting in invalid outputs. This can be mitigated by pruning such paths through lightweight syntax and schema validation before full execution.

\textbf{Reward evaluation inconsistency (17.1\%)} Although the correct query may be generated, the node-value $V(\cdot)$ estimates during the back-propagation phase often deviate from the optimal values, causing the model to select paths that receive high rewards despite containing logical errors. This issue primarily arises from imperfect heuristic reward estimation during search. Potential solutions include developing more reliable reward signals or improving trajectory-value estimation.

\textbf{Correct path not discovered (26.2\%)} The MCTS fails to cover the correct “golden” query, especially in complex cases involving joins, subqueries, or nested conditions. This is often due to limited depth or insufficient structural priors. Increasing search depth and enriching training data can help mitigate the issue.

\textbf{Other Errors (5.0\%)} This category includes infrequent issues that do not fall into the main error types, such as schema misinterpretations and inaccuracies in the gold query annotations within the training data. Although each individual occurrence is rare, these errors collectively contribute to overall performance degradation. Addressing these issues requires improving dataset quality, refining schema linking mechanisms, and enhancing the model's robustness to noisy data.

\begin{figure}[h] 
    \centering
    \includegraphics[width=\linewidth]{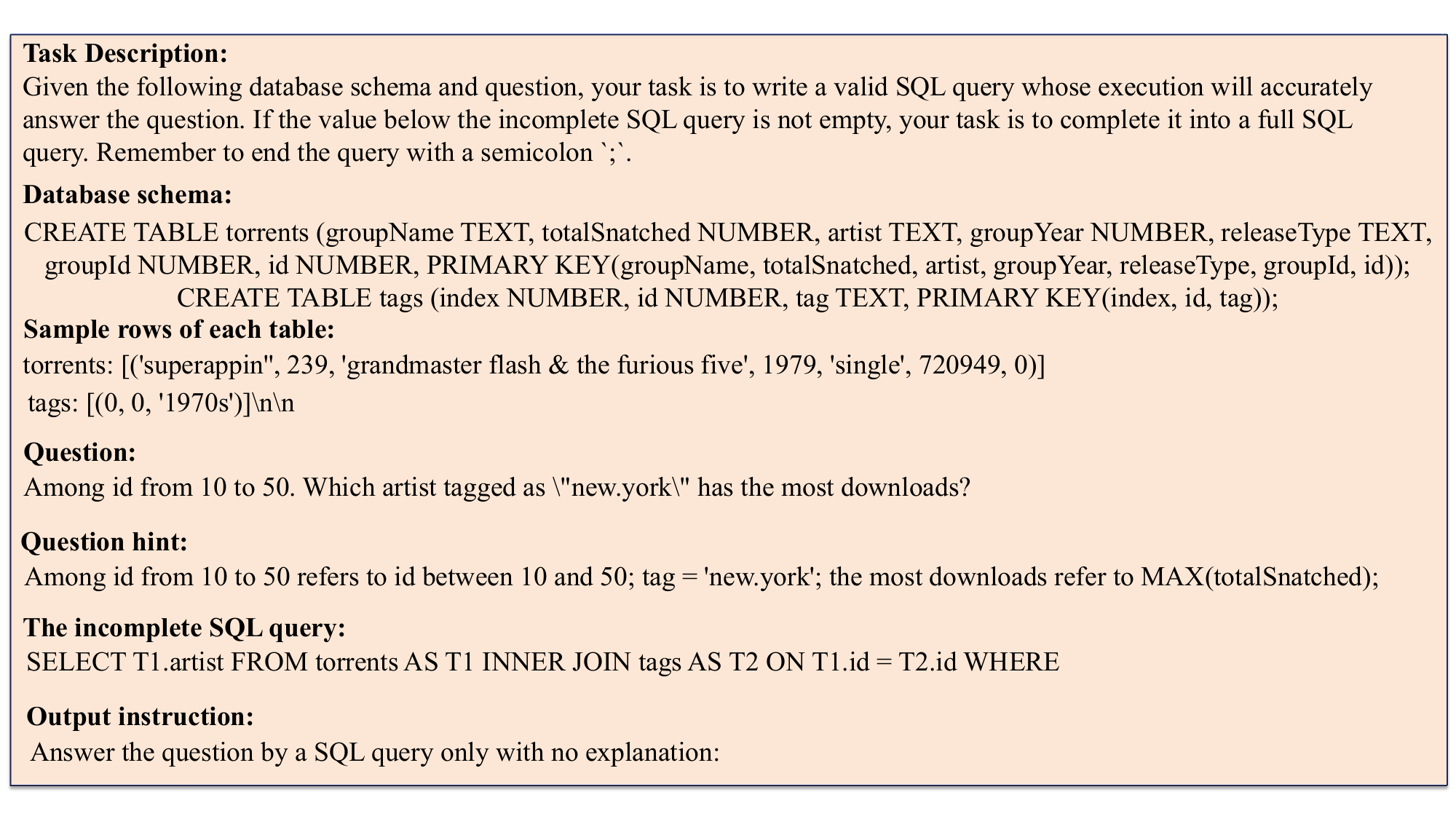}  
    \caption{The prompt template for Text-to-SQL in the fine-tuning phase.}
    \label{fig:prompt}
    \Description[The structure of a Text2SQL fine-tuning prompt]{
        The figure illustrates the template used to create prompts for fine-tuning a Text2SQL model. It likely shows a structured text format containing a natural language instruction or question, the corresponding ground truth SQL query, and potentially special tokens or separators used to format the input for the language model.
    }
    \vspace{-6mm}
\end{figure}

\subsection{Prompt Templates}
In this section, we provide the task prompts used in SQL-o1 for Schema-Aware and Progressive SQL Generation. Since the Spider and BIRD datasets are quite similar, we only present the BIRD task prompts, as shown in Figure~\ref{fig:prompt}.

\section{Further Discussions}
\subsection{Why SQL-o1 Prefers MCTS to Sequence-Level RL Objectives (GRPO / PPO / DPO)?}
In contrast to open-ended reasoning tasks, SQL generation exhibits two task-specific properties that make it particularly suitable for a \emph{step-level} search paradigm.

\paragraph{Block-level, schema-constrained action space.}%
At each step, SQL generation selects from a finite set of semantic blocks—\verb|SELECT|, \verb|WHERE|, \verb|JOIN|—or table–column pairs, and the search depth rarely exceeds ten. MCTS leverages model confidence and lightweight validity checks to prune nodes early, keeping exploration focused on promising candidates; GRPO/PPO must still back-propagate gradients over vocabularies with thousands of tokens, resulting in poor sample efficiency.

\paragraph{Dense and immediate rewards.}%
Partial SQL prefixes provide structured intermediate states that can be evaluated using model confidence and lightweight validity checks, while complete rollouts additionally receive execution-agreement feedback. MCTS propagates these signals through the search tree, reducing reliance on a single terminal decision.

\paragraph{Negative-sample bottleneck.}%
In our experiments we adopted Direct Preference Optimization (DPO) as an RL baseline to guide SQL exploration. However, DPO is only effective when high-quality negative samples are available, such as query pairs that are structurally different but semantically similar. It struggles to discern such differences in structured outputs and therefore ended up performing worse than supervised fine-tuning (SFT) that uses positive samples only.

In summary, considering the controllable action space, dense reward signals, and need for interpretability, MCTS is particularly well suited to the Text-to-SQL setting considered in this work.

\subsection{SQL-o1 Outperforms Beam Search and Best-of-N Sampling in Text-to-SQL.}
Beam Search and Best-of-N sampling are among the most widely used decoding strategies in sequence generation tasks, and are typically employed as baseline inference methods for generating structured queries. 
However, both approaches exhibit significant limitations. 
Beam Search preserves only the top-k partial sequences at each decoding step, lacking the capacity for global semantic optimization. 
Best-of-N sampling generates multiple complete queries through independent sampling, but its unstructured exploration often leads to invalid or semantically inconsistent outputs. 
More critically, neither method is able to incorporate execution feedback to revise the search trajectory dynamically, often resulting in locally optimal or structurally incomplete predictions.

In contrast, SQL-o1 adopts an MCTS-based search mechanism guided by model-based self-reward together with online execution feedback.
It dynamically evaluates the utility of candidate paths and updates the search tree through value back-propagation.
Compared to the deterministic expansion in Beam Search and the unconstrained sampling of Best-of-N, SQL-o1 introduces structural interpretability and reward-guided exploration, enabling the discovery of semantically correct but low-probability queries that standard methods often overlook.

\subsection{Role of Log-Likelihood-Based Self-Reward} Although the self-reward in Equation~\eqref{eq:reward} is derived from the model's log-likelihood, SQL-o1 does not use it as an independent correctness criterion or final decision rule. Instead, it serves as a local preference heuristic within MCTS. UCT-based selection introduces explicit exploration, while Equation~\eqref{eq:Q_value} further incorporates online execution agreement from repeated rollouts. The resulting trajectory values are back-propagated through the search tree, and confidence-based dynamic pruning controls subsequent exploration. Therefore, the final SQL is determined by the overall search-and-feedback process rather than by likelihood alone. Nevertheless, model confidence remains heuristic and does not guarantee semantic correctness.

\subsection{Software Engineering Implications} SQL-o1 provides a program-synthesis component for natural-language database interfaces. By constructing partial SQL programs under schema constraints and evaluating candidates through execution feedback, the framework helps reduce the risk of incorrect joins, filters, or aggregations affecting downstream application behavior. The MCTS search maintains intermediate states and alternative trajectories, making the generation process easier to inspect and supporting the validation and debugging of generated queries, although its impact on developer productivity requires empirical evaluation. Deployment exposes an accuracy--cost trade-off controlled through rollout count, search depth, and pruning threshold. More aggressive pruning can reduce inference time at the cost of accuracy, while additional rollouts generally improve accuracy with greater overhead. We report token consumption and inference cost across configurations to guide deployment under varying latency and resource constraints.


\subsection{Limitations and Future Work} 
\paragraph{Annotated Data Scarcity.} Text-to-SQL lacks step-level annotations, complicating training and generalization. We will explore obtaining intermediate supervision without manual annotation.
\paragraph{Reward and Search Limitations.} SQL-o1 evaluates trajectories using model confidence and execution agreement among rollouts, as ground-truth SQL is unavailable during inference. Without direct feedback from the gold SQL, reward calibration therefore relies on these indirect signals. The fixed search depth and rollout budget also constrain coverage, potentially excluding correct solutions to complex queries. Improved reward estimation and adaptive search allocation remain important directions for improvement. 
\paragraph{Execution Dependence and Computation Cost.} MCTS incurs substantial inference overhead. SQL-o1 also requires database access for execution feedback, which may be costly or unavailable in some deployments. Beyond dynamic pruning, we will investigate adaptive scheduling and lightweight execution simulation to reduce both search cost and database dependence.

\section{Conclusion}
We introduce SQL-o1, a self-reward-driven heuristic search framework that enhances the reasoning capabilities of large language models (LLMs) in Text-to-SQL synthesis for database-backed applications. By combining Monte Carlo Tree Search (MCTS) with schema-aware training, SQL-o1 improves SQL generation accuracy, achieving a 10.8-point gain on BIRD over the same-backbone baseline. It also demonstrates few-shot generalization and cross-model transferability.

\begin{acks}
This work is supported by the
\grantsponsor{nkrdp}{National Key Research and Development Program of China}{https://doi.org/10.13039/501100012166}
under Grant~\grantnum{nkrdp}{2024YFC3308500},
\grantsponsor{bjnsf}{Beijing Municipal Natural Science Foundation}{https://doi.org/10.13039/501100005089}
under Grant~\grantnum{bjnsf}{L251042},
\grantsponsor{nsfc}{National Natural Science Foundation of China}{https://doi.org/10.13039/501100001809}
under Grant~\grantnum{nsfc}{62406036} and also sponsored by the
\grantsponsor{sklnst}{State Key Laboratory of Networking and Switching Technology}{https://sklnst.bupt.edu.cn/}
under Grant~\grantnum{sklnst}{NST20250110}.
\end{acks}



\bibliographystyle{ACM-Reference-Format}
\bibliography{tosem}

\end{document}